\documentclass[aps, prl, twocolumn, showpacs, floatfix,10pt,superscriptaddress]{revtex4-2}
\usepackage[T1]{fontenc}
\usepackage{amsmath}
\usepackage{amsfonts}
\usepackage{amsbsy}
\usepackage{amssymb}
\usepackage{graphicx}
\usepackage{textcomp}
\usepackage[caption=false,singlelinecheck=false]{subfig}
\usepackage{xcolor}
\usepackage{mathrsfs}
\usepackage{mathtools}
\usepackage{bm}
\usepackage{gensymb}
\usepackage{braket,wasysym}
\usepackage[colorlinks,linkcolor=blue,citecolor=red,filecolor=magenta,urlcolor=red,breaklinks]{hyperref}
\usepackage[bottom]{footmisc}
\usepackage{verbatim}

\hypersetup{colorlinks=true, urlcolor=blue, citecolor=red, pdfborder={0 0 0}}
\usepackage{breakurl}
\usepackage{natbib}

\allowdisplaybreaks

\newcommand{\tcm}[1]{\textcolor{magenta}{#1}}

\normalfont
\def\be{\begin{equation}}
	\def\ee{\end{equation}}
\def\bea{\begin{eqnarray}}
	\def\eea{\end{eqnarray}}

\begin{document}
\title{Entanglement switching via mobility edges in a quasiperiodic chain}
\author{YouYoung Joung}
\email{young_sl.physics@kaist.ac.kr}
\affiliation{Department of Physics, Korea Advanced Institute of Science and Technology, Daejeon, 34141, Korea}
\author{Junmo Jeon}
\email{junmo1996@kaist.ac.kr}
\affiliation{Department of Physics, Korea Advanced Institute of Science and Technology, Daejeon, 34141, Korea}
\author{SungBin Lee}
\email{sungbin@kaist.ac.kr}
\affiliation{Department of Physics, Korea Advanced Institute of Science and Technology, Daejeon, 34141, Korea}

\date{\today}
\begin{abstract}
We propose quasiperiodic chains with tunable mobility edge physics, as a promising platform for engineering long-range quantum entanglement.
Using the generalized Aubry-André model, we show that the mobility edges play a key role in manipulating long-range indirect interactions in these systems. Near the mobility edge, critical states exhibit unexpectedly strong correlations between sites that share similar local structures, regardless of their spatial separation. Remarkably, by tuning the mobility edge across the Fermi level, one can induce both adiabatic transport and abrupt switching of entanglement between distant sites.
These results highlight the potential of aperiodic structures for controlling nonlocal quantum correlations, opening new avenues for entanglement-based applications in quasiperiodic systems. 
\end{abstract}


\maketitle

\textit{\tcm{Introduction---}} The generation and control of entanglement across far distant subsystems are central tasks of quantum technology \cite{RevModPhys.81.865, audretsch2007entangled}, underpinning applications from quantum computing \cite{PhysRevA.62.062311, PhysRevA.89.022317} to secure communication \cite{PhysRevA.65.032302, kimble2008quantum, RevModPhys.74.145} and metrology \cite{PhysRevA.54.R4649, PhysRevLett.96.010401}. In conventional solid-state or optical platforms, however, sustaining coherent interactions over long distances is fundamentally limited by decoherence \cite{PhysRevLett.77.3240, RevModPhys.73.565, haroche1998entanglement} and by the rapid spatial decay of coupling strengths.
To overcome these limitations, significant recent efforts have focused on various approaches, including robust topological structures \cite{tantivasadakarn2024long, hetenyi2022long, tang2024topologically, wang2020topologically, mittal2018topological}, atom-level engineering via optical tweezers \cite{omran2019generation, dhordjevic2021entanglement, shaw2025erasure, hartung2024quantum}, measurement-based quantum circuits \cite{iqbal2024topological, baumer2025measurement, baumer2024efficient, lu2022measurement}, the use of entangled photon beams\cite{sangouard2011quantum, PhysRevA.105.062454, krutyanskiy2023entanglement, qiu2024efficient}, and so on \cite{burkard2023semiconductor, lei2023quantum, wei2022towards}. 
Thus, the stable generation and precise manipulation of quantum entanglement over extended distances represent both a central goal and a major challenge in the development of quantum communication technologies \cite{PhysRevA.105.062454, PhysRevLett.126.010503, ursin2007entanglement, yuan2010entangled, duan2001long}.

Quasiperiodic systems, with their long-range order in the absence of translational symmetry, have recently attracted interest as possible candidates for enabling stable long-range entanglement\cite{PhysRevB.111.L020411}. 
In metallic quasiperiodic systems, the presence of critical states, neither fully localized nor extended, gives rise to wavefunctions with power-law decay and fractal structure \cite{wilkinson1984critical, PhysRevLett.51.1198, aubry1980analyticity}. These critical states have been shown to mediate anomalously strong indirect interactions between spatially separated localized spins, potentially enabling long-range entanglement that remains robust against thermal fluctuations \cite{PhysRevB.111.L020411}. However, a key challenge in implementing this scheme lies in the fact that the nature of such indirect interactions is largely governed by the position of the Fermi level \cite{shklovskii2013electronic}. While the Fermi level can, in principle, be tuned through doping, achieving precise and systematic control over it remains technically demanding in most metallic systems
\cite{shklovskii2013electronic}.

In this work, we propose quasiperiodic systems with tunable mobility edges as a promising platform for realizing long-range entanglement and enabling controllable entanglement switching simultaneously. Specifically, we focus on the experimentally accessible generalized Aubry–André (GAA) model \cite{PhysRevLett.114.146601, PhysRevLett.104.070601, PhysRevLett.126.040603, PhysRevLett.129.103401}, where tuning the mobility edge plays a crucial role in overcoming limitations addressed above, by allowing critical states to mediate strong long-range spin–spin interactions. Furthermore, we demonstrate two distinct modes of entanglement control—adiabatic transfer and sharp on–off switching—both achievable without modifying the filling fraction.
Our findings establish quasiperiodic systems, characterized by intrinsic pattern similarity, as a powerful and experimentally controllable platform for programmable entanglement distribution, laying the groundwork for an extended perspective on scalable quantum architectures.

\textit{\tcm{Generalized Aubry-Andr\'e model with magnetic impurities}---} 
Let us consider the tight-binding model of 1D quasiperiodic potential with localized magnetic impurities. The Hamiltonian is written as,
\begin{align}
    \label{H}
    &H=-t\sum_{i,\sigma}(c^\dagger_{i+1,\sigma}c_{i,\sigma}+\mathrm{h.c.})+\sum_{i,\sigma} V_in_{i,\sigma}+J_K\sum_i\vec{S}_i\cdot\vec{s}_i,
\end{align}
where $c_{i,\sigma}^\dagger$ and $c_{i,\sigma}$ are creation and annihilation operators of tight-binding electron at site $i$ with spin $\sigma$, where $0\leq i <N$. $n_{i,\sigma}=c_{i,\sigma}^\dagger c_{i,\sigma}$. $t$ is the uniform hopping integral. We set $t=1$. $J_K$ is the local exchange coupling between effective spin-1/2 impurity, $\vec{S}_i$ and tight-binding electron spin $\vec{s}_i=c_{i,\sigma}^\dagger \vec{\sigma}_{\sigma,\sigma'} c_{i,\sigma'}/2$. $V_i$ is the quasiperiodic potential.

Through this paper, we consider $V_i$ as generalized Aubry-Andr\'e (GAA) potential given by,
\begin{align}
\label{GAApotential}
&V_{i}=\lambda\frac{\cos(2\pi qi+\phi)}{1-\alpha \cos(2\pi qi+\phi)}.
\end{align}
 Here, $\phi$ is an arbitrary phase shift, set as $\phi=0$ without loss of generality. This potential becomes quasiperiodic for arbitrary irrational number $q$ \cite{PhysRevLett.114.146601}. For concrete argument, we set $q=(\sqrt{5}-1)/2$. 
 The parameter $-1<\alpha<1$ and $\lambda$, which are experimentally tunable in various platforms including optical lattice in terms of digital micromirror devices \cite{qiu2020precise, lukin2019probing} or multifrequency lasers \cite{PhysRevLett.126.040603, PhysRevLett.129.103401}, determine the potential profile and localization characteristics in the spectrum of tight-binding Hamiltonian. In detail, nonzero $\alpha$ has nontrivial mobility edge (ME), $E_{\mathrm{ME}}$ in the spectrum as \cite{PhysRevLett.114.146601}
\begin{align}
    \label{ME}
    \alpha E_{\mathrm{ME}} =\mathrm{sgn}(\lambda)(2\vert t\vert - \vert\lambda\vert).
\end{align}
Near this ME, the states exhibit neither localized nor extended but critical behaviors. While, $\alpha=0$ corresponds to the Aubry-Andr\'e model, which has self-dual point at $\vert\lambda\vert/t=2$, independent to energy, leading to the absence of ME \cite{aubry1980analyticity}. Such a shift in $\alpha$  can be experimentally realized by precise site-by-site manipulation of the on-site energy. A digital micromirror devices, composed of individually controllable pixels of mirrors, for instance, can project tailored light patterns onto a cold atom system, enabling the creation of arbitrarily shaped optical potentials. \cite{qiu2020precise, gauthier2016direct} Thus, the electronic characteristics with potential function in Eq.\eqref{GAApotential} is highly tunable in experiments.

When $\vert J_K\vert\ll 1$, one can consider the effective long-range interactions between spatially distant impurity spins. In detail, by integrating out  the tight-binding electrons, an effective exchange Hamiltonian for the impurity spins, is derived, 
$\mathcal{H}=\sum_{i\neq j} J_{ij}\vec{S}_i\cdot\vec{S}_j$. Here, the effective indirect long-range interaction, $J_{ij}$ between $i$ and $j$ sites is given by \cite{PhysRev.96.99, PhysRevB.36.3948}
\begin{align}
\label{RKKY2}
J_{i,j}=\frac{J_K^2}{4}\sum_{m,n,E_m\neq E_n}&\mathcal{R}[\psi_m(i)\psi_m(j)^*\psi_n(j)\psi_n(i)^*] \\ &\times \frac{n_F(E_n-E_F)-n_F(E_m-E_F)}{E_n-E_m}.\nonumber
\end{align}
We set $\hbar=k_B=1$. $E_F$ is the Fermi energy. $\mathcal{R}[x]$ is real part of $x$, respectively. $n,m$ are indices of the eigenstates, and $\psi_{n}(p)=\langle p \vert n\rangle$ is the wave function of the energy eigenstate $\vert{n}\rangle$ whose energy is $E_n$ at the $p$ site. $n_F(x)=(1+\exp(x/T))^{-1}$ is the Fermi-Dirac distribution function, where $T$ is temperature. In periodic system which premises a Bloch type wavefunction, $J_{ij}$ shows $2k_F$ oscillation with a power-law decay of $\sim \vert i-j\vert^{-1}$ scaling in one dimension \cite{PhysRevB.36.3948}. However, in the case of GAA model with irrational $q$, the direct application of such conventional behaviors of $J_{ij}$ is not valid due to the absence of periodicity. This allows us to explore the unconventional long-range interactions in this model as we will show. 

\begin{figure}[htb]
    \centering
	\includegraphics[width=1\linewidth]{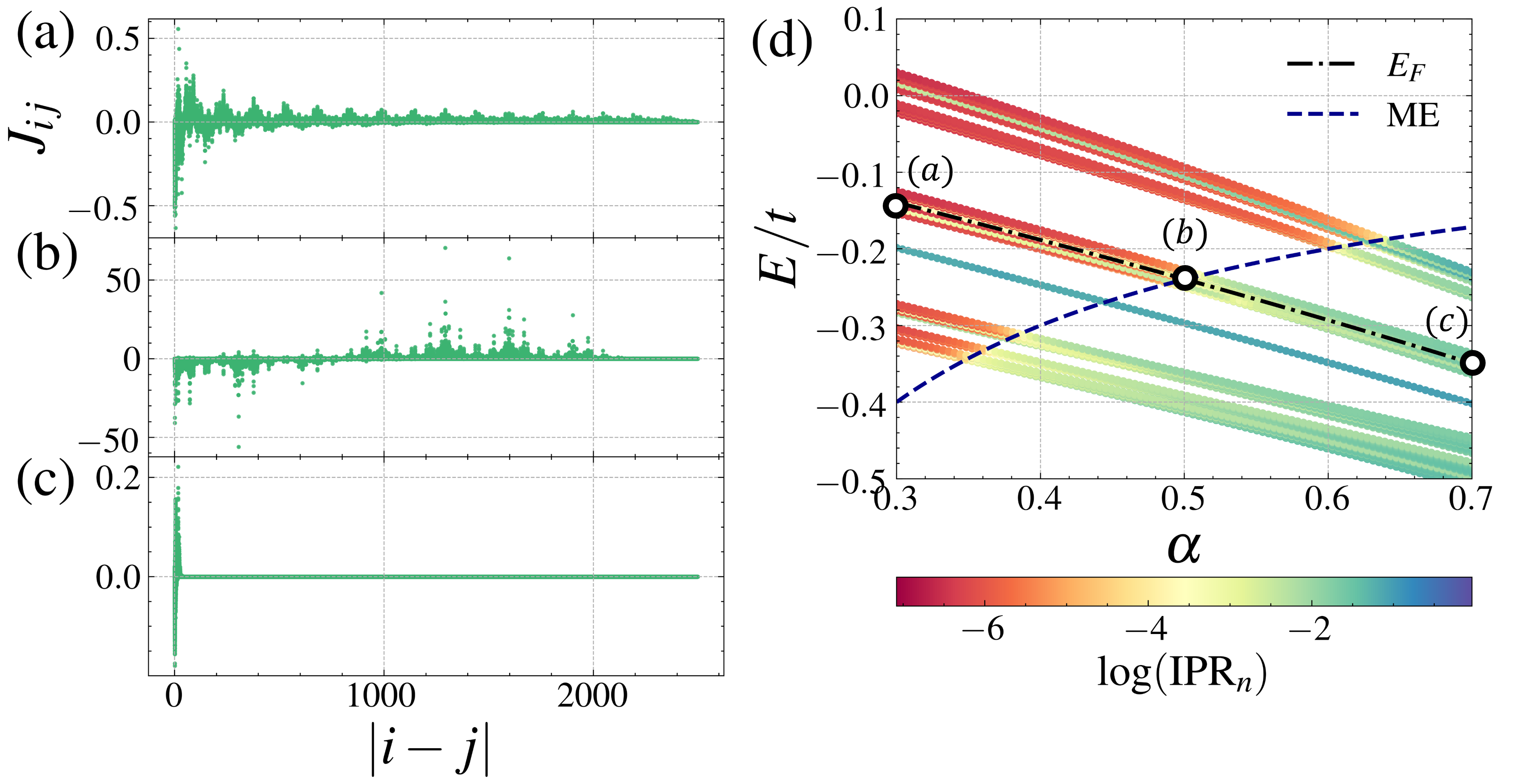}
        \caption{Three distinct properties under the shift of the ME via $\alpha$ modulation. (a-c) The indirect long-range interaction strength $J_{ij}$ for $J_{K}/t=1$ as a function of site distance $\vert i-j\vert$, in the GAA model with parameters $\lambda/t=-1.88$ and $\phi=0$ for (a) $\alpha=0.3$, (b) $\alpha=0.5$, and (c) $\alpha=0.7$. The Fermi energies for each $\alpha$ values are marked in (d). Here, the system is half-filled. (d) The energy spectrum as the function of $\alpha$. The colors represent the logarithm of inverse participation ratio, $\mathrm{IPR}_n$ of each eigenstate. The mobility edge (ME) and the Fermi energy ($E_F$) of half filling are drawn as dotted and dashed curves, respectively. $N=2500$ and $T=0$.}
    \label{fig:3alpha}
\end{figure}

\textit{\tcm{Tunable long-range interaction}---} 
In general, one conventional way to modify the characteristics of the long-range interaction described by  Eq.\eqref{RKKY2} is by shifting the Fermi level. 
However, in metals, the Fermi level is determined by the atomic composition and is therefore difficult to adjust. Techniques such as doping are also limited in their ability to induce significant changes \cite{shklovskii2013electronic}. On the other hand, the GAA model, which can be realized in experiments using the optical lattices \cite{PhysRevLett.126.040603, PhysRevLett.129.103401, PhysRevLett.104.070601}, allows us to control ME in terms of experimentally tunable parameters, $\alpha$.
This enables us to control the indirect spin-spin interaction mediated by itinerant electrons in a quasiperiodic chain, while fixing the filling fraction. In detail, as the parameter $\alpha$ in Eq.\eqref{GAApotential} varies, the mobility edge changes according to Eq.\eqref{ME} unless $\vert \lambda\vert= 2\vert t\vert$. Thus, for given filling fraction, the localization characteristics of the states near the Fermi level changes with $\alpha$. Consequently, the characteristics of the indirect interaction could be drastically manipulated in terms of $\alpha$ as we will show.

Figures~\ref{fig:3alpha} (a-c) display three distinct cases of indirect spin-spin interaction at the half-filling for different $\alpha$. Note that Figures~\ref{fig:3alpha} (a) and (c) exhibit conventional spin-spin interactions found in metallic and insulating regime, respectively. Specifically, $J_{ij}$ shown in Fig.\ref{fig:3alpha} (a) and (c) are power-law and exponentially decaying as the function of physical distance, $\vert i-j\vert$, respectively. In contrast, Figure~\ref{fig:3alpha} (b) illustrates a striking enhancement of the interaction strength at selected long-range site pairs. In particular, the strongest coupling exceeds that in panels (a) and (c). More surprisingly, the strong couplings emerge between anomalously long-distant sites, for instance $i=646$ and $j=1938$. We emphasize that the indirect interaction, $J_{ij}$ shown in Fig.\ref{fig:3alpha} (b), depends on both the positions of $i$ and $j$, not just the distance between them.

To figure out $\alpha$-dependent long-range interactions, we examine the localization characteristics near the Fermi level in terms of inverse participation ratio (IPR) given by $\mathrm{IPR}(\psi)=\sum_{i=1}^N\vert\psi(i)\vert^4$, where $\psi$ is the normalized state \cite{bauer1990correlation, PhysRevLett.114.146601}. Note that IPR of the extended state falls off as the system size increases, while the IPR of the localized state remains constant. The intermediate values of IPR indicates the critical states that are neither fully localized nor extended but concentrated on some selective positions with power-law scaling behavior.
Such critical states appear within the tiny energy window around the mobility edge given by Eq.\eqref{ME}. Since the energy of the mobility edge, $E_{\mathrm{ME}}$ depends on $\alpha$ as given in Eq.\eqref{ME}, the localization characteristics of the states near the Fermi level also change as a function of $\alpha$, even though the filling fraction is fixed.

Figure~\ref{fig:3alpha} (d) exhibits relative positions between $E_\text{ME}$ and Fermi energy on the spectrum of tight-binding Hamiltonian as the function of $\alpha$. When the ME is lower (higher) than $E_F$, the states near the Fermi level is extended (localized). While, when $E_F\approx E_{\mathrm{ME}}$, the states near the Fermi level are critical states.
Note that when $\alpha>0$, $E_F$ increases with $\alpha$ because $\alpha$ enhances average on-site potential energies described in Eq.\eqref{GAApotential}. In contrast, Eq.\eqref{ME} indicates that $E_{\mathrm{ME}}$ decreases as $\alpha$ increases when $\lambda<0$ and $2\vert t\vert >\vert \lambda\vert$. Thus, at $\alpha=\alpha_c$, where the Fermi level coincides with ME, the critical states play a crucial role in long-range coupling.

\begin{figure}[t!]
    \centering
    \includegraphics[width=1\linewidth]{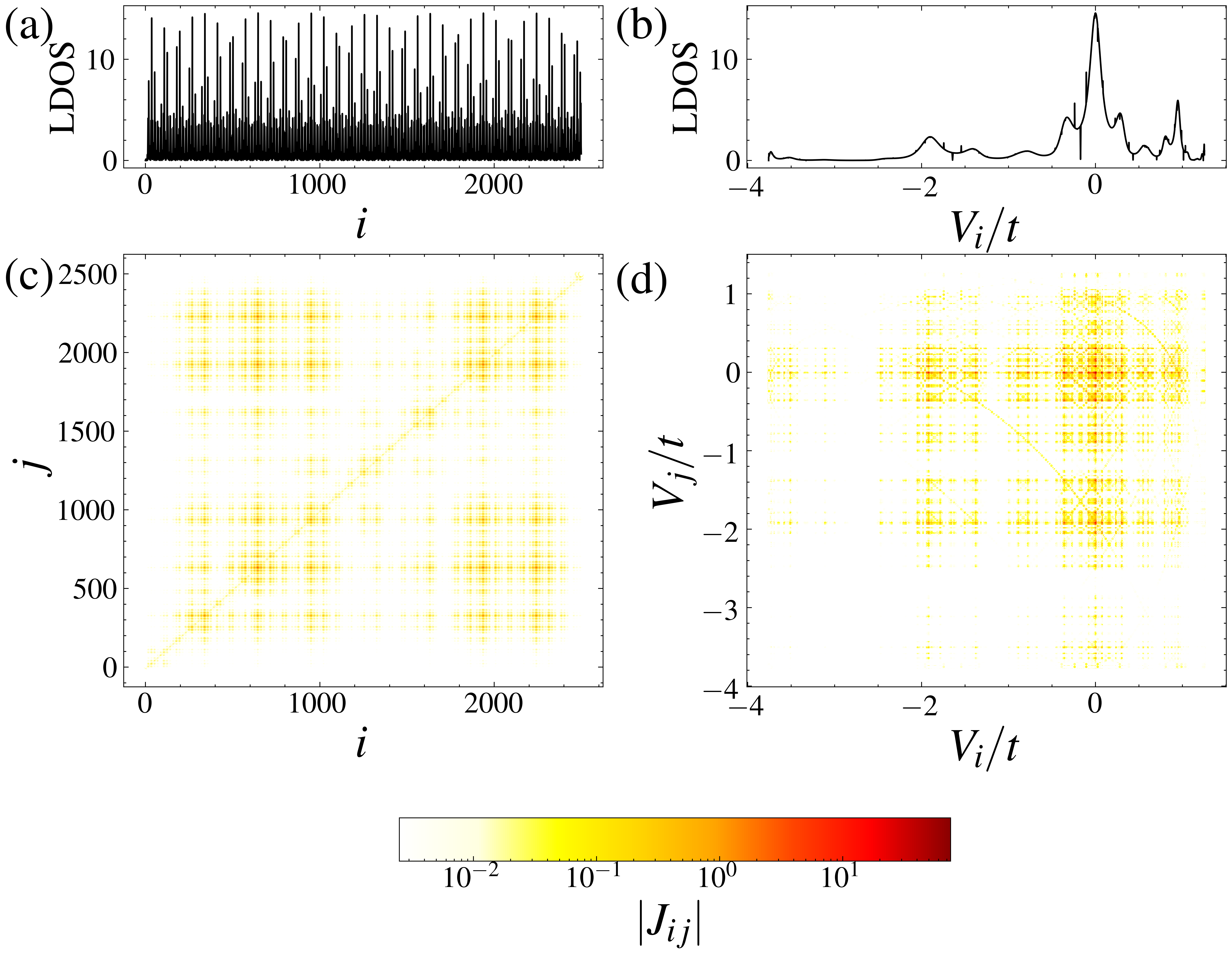}
       \caption{Pattern-selective interaction in the generalized Aubry--Andr\'e (GAA) model. (a-b) Local density of states of GAA model at the half filling Fermi energy for each site as a function of (a) site indices (b) on-site potential $V_{i}$. (c-d) The absolute value of the interaction strength $|J_{ij}|$ when $J_{K}=1$ at the zero temperature limit as a function of (c) site indices (d) on-site potential energies. Here, $\alpha=0.5$, $\lambda=-1.88$.}
    \label{fig:Internal}
    
\end{figure}

\textit{\tcm{Pattern-selective interaction---}} Let us focus on the case of anomalously enhanced long-range interactions originated from the critical states around the Fermi level. Despite the anomalous long-range interactions $J_{ij}$ exemplified in Fig.\ref{fig:3alpha} (b) depend on both site indices $i$ and $j$ rather than their simple spatial distance (see Fig.\ref{fig:Internal} (c) for general pairs of sites), we assert that the pairs of strongly interacting sites could be selectively chosen by the similarity of the local patterns given by the quasiperiodic potential Eq.\eqref{GAApotential}. To clarify our claim, we investigate the local density of states (LDOS) at the Fermi energy given by $\sum_n\vert \psi_n(i)\vert^2\delta (E_F-E_n)$ \cite{niimi2006scanning, chicanne2002imaging} as the function of the on-site potential energy, $V_i$. Note that Eq.\eqref{GAApotential} is a quasiperiodic function, and hence $V_i\neq V_j$ for every $i\neq j$.

Figures~\ref{fig:Internal} (a) and (b) exhibit the LDOS at the Fermi energy as the function of position and on-site potential energy, respectively. Here, $E_{\mathrm{ME}}\approx E_F$. The rearrangement reveals the hidden structure of LDOS concentration around $V_{i} = 0$ (see Fig.\ref{fig:Internal} (a-b)). This is because the critical states near the Fermi level mostly concentrated on the sites of $V_{i}\approx0$. Accordingly, the long-range interaction also follows the trend (see Fig.\ref{fig:Internal}(c-d)). Thus, the local structure determines the unconventional interaction.

Such pattern-selective interaction—whose strength is proportional to the similarity of the local electron Hamiltonian—is a universal feature of spin-spin interactions mediated by critical states in generic incommensurate systems. In the GAA model, this similarity is captured by having similar on-site potential values, whereas in more general incommensurate and quasiperiodic systems, it originates from local structural patterns defined by kinetic terms or potential distributions \cite{PhysRevB.111.L020411}. Unlike periodic or amorphous systems, quasiperiodic incommensurate systems feature locally repeating patterns without exact global repetition. As a result, critical states near the Fermi level tend to strongly occupy sites with similar local environments across wide regions, significantly enhancing the local density of states and long-range interaction strength between these sites, regardless of their spatial separation.

\begin{figure*}[htb]
    \centering
    \includegraphics[width=1\linewidth]{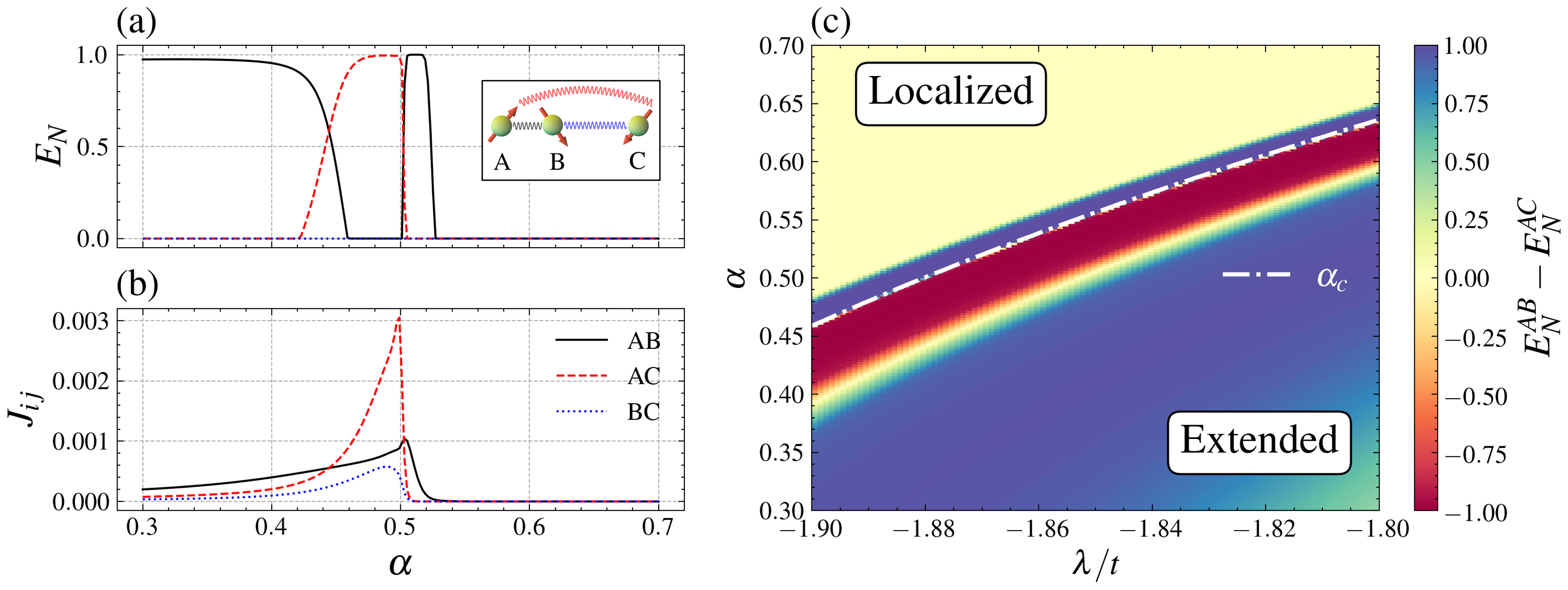}
    \caption {Entanglement transfer and switching between the short range pair A, B $(646, 879)$ and the long range pair A, C ($646, 1938$). (a) Schematic illustration of three spin interacting system and plot of logarithmic negativity as a function of $\alpha$ with $\lambda=-1.88$ fixed. The critical point, where the entanglement switching happens is $\alpha_{c}\approx0.501$. (b) Interaction strength $J_{ij}$ in the same condition, between all 3 possible spin pairs AB, BC and AC in solid, dotted, and dashed curve respectively. (c) The landscape of difference of logarithmic negativity of short range (AB pair) and long range (AC pair) as the function of $\alpha$ and $\lambda$. The white dashed dot curve is drawn to emphasize the critical value $\alpha_{c}$, where the entanglement switching between two spin pairs occurs as the function of $\lambda$. Here, we present the results obtained at fixed half-filling, $T/t=10^{-4}$ and $J_{K}/t=0.05$.}
    \label{fig:Monogamy}
\end{figure*}

\textit{\tcm{Entanglement transfer and switching---}} Experimentally tunable pattern-selective long-range interactions bring a  notable advancement in controlling long-range entanglement. Remind that long-range entanglement between distant qubits underpins many emerging quantum technologies, including quantum computing and communication \cite{PhysRevA.62.062311, PhysRevA.65.032302, gisin2007quantum}, but is notoriously difficult to achieve and control at finite temperatures \cite{PhysRevA.77.012117}. One of the reason is because typical long-range interactions decay rapidly with real-space distance \cite{PhysRev.96.99}, requiring thermodynamically many intermediary qubits. This makes the system highly vulnerable to decoherence. However, in incommensurate systems, the spin interactions depend not on physical distance but on inherent local pattern similarity, leading to strong entanglement between a few widely separated spins \cite{jeon2025immortal}.

Furthermore, we emphasize that the high experimental tunability of the GAA model allows on-demand switching and transfer of entanglement between widely separated spins in terms of the systematic parameters such as $\alpha$ and $\lambda$ without extra doping.        
To be more specific, let us show how systematic parameters in the GAA model lead to the unconventional control of long-distance entanglement while keeping the filling-fraction.
To demonstrate such high-controllability of the long-distance entanglement driven by $\alpha$ and $\lambda$, we consider three localized spin-1/2 impurities placed on $i,j$ and $k$ sites, respectively, where the differences in their on-site potentials satisfy $\vert V_i-V_j\vert^2+\vert V_j-V_k\vert^2+\vert V_k-V_i\vert^2\ll 1$, indicating that all three sites have nearly identical local environments. Additionally, the potential differences obey the ordering $\vert V_k-V_i\vert<\vert V_i-V_j\vert<\vert V_j-V_k\vert$.
We denote the spins at sites $i$, $j$, and $k$ as A, B, and C, respectively. Let us assume that the interactions among these spins are all antiferromagnetic within the given range of $\alpha$ (see Fig.\ref{fig:Monogamy} (b)). As a concrete example, we consider $(i,j,k)=(646,879,1938)$.

Given $\vert \lambda\vert \neq 2t$, we note two remarkable features of the long-range interactions. First, as $\alpha$ approaches the critical value $\alpha_c$ from $\alpha<\alpha_c$, $J_{ki}$ increases significantly compared to $J_{ij}$ and $J_{jk}$, since $\vert V_j-V_k\vert>\vert V_i-V_j\vert>\vert V_k-V_i\vert$. In contrary, as $\alpha$ exceeds the critical point ($\alpha\ge \alpha_c$), $J_{ki}$ suddenly drops. Note that for $\alpha>\alpha_c$, localization near the Fermi level leads to an exponential decay of interaction strength with distance.

Figure~\ref{fig:Monogamy} (a) shows unconventional control of the entanglement driven by anomalous behaviors of long-range interactions as the function of $\alpha$ at finite temperature. Here, we adopt the logarithmic negativity $E_{N}^{AB (AC)}=\log_{2}(\lVert (\rho_{C(B)})^{\Gamma_A}\rVert_{1})$ as an entanglement measure for the reduced density matrix $\rho_{C(B)}=\mathrm{Tr}_{C(B)}\rho$ , where $\Gamma_{A}$ indicates the partial transpose with respect to $A$ and $\lVert X \rVert_{1}=\mathrm{Tr}\sqrt{ X^{\dagger}X }$ is the trace norm \cite{PhysRevA.65.032314, horodecki2001entanglement}. The entanglement between each spin pair reflects the relative strength of long-range interactions. In detail, when $J_{ki}>J_{ij}$, entanglement between spins A and C is larger than that between A and B. This is because of monogamy of entanglement \cite{koashi2004monogamy}. Since spins B and C compete for entanglement with spin A, an increase in entanglement between A and B (say AB pair) necessarily leads to a decrease in entanglement between A and C (say AC pair). Consequently, following the anomalous behavior of the long-range interactions shown in Fig.~\ref{fig:Monogamy} (b), the entanglement of AB pair and AC pair exhibit two kinds of useful changes as the function of $\alpha$. Specifically, as $\alpha<\alpha_c$ increases, the entanglement is smoothly transferred from AB pair to AC pair, while as $\alpha>\alpha_c$, the entanglement of AB (AC) pair is suddenly turned-on (off), respectively. This is because as $\alpha$ exceeds $\alpha_c$, localization causes distance-dependent decay, making $J_{ki}$ drop faster than $J_{ij}$.
Thus, both the transfer and switching of entanglement,typically difficult or even unachievable in conventional systems \cite{hartung2024quantum, baumer2025measurement}, 
become naturally accessible through the tunable parameter 
$\alpha$ and the intrinsic quasiperiodic structure of the GAA model.
Note that non of the spin pairs exhibit finite-temperature entanglement in the fully localized regime $\alpha\gg \alpha_{c}$ by their spatial separations. The nonzero entanglement arises only at the extended regime $\alpha<\alpha_{c}$ and critical regime $\alpha \approx \alpha_{c}$, reflecting the pattern selective anomalous interaction.

We emphasize that the entanglement transfer and switching near the critical point $\alpha_c(\lambda)$ reflect a generic, tunable mechanism applicable to arbitrary $\lambda$, not a $\lambda$-specific feature. Figure~\ref{fig:Monogamy} (c) illustrates the difference in logarithmic negativity between the AB and AC pairs in the vicinity of $\alpha_c(\lambda)$ for general $\lambda$. Given $\vert\lambda\vert\neq 2t$, as $\alpha$ increases toward $\alpha_c(\lambda)$, the logarithmic negativity of the AC pair ($E_N^\mathrm{AC}$) rises smoothly from 0 to its maximum value of 1, while $E_N$ of the AB pair ($E_N^\mathrm{AB}$) decreases from 1 to 0. This is the long-distance entanglement transfer. Meanwhile, as $\alpha$ slightly exceeds $\alpha_c(\lambda)$ and the Fermi level starts to enter the localized regime, $E_n^\mathrm{AB}$ and $E_n^\mathrm{AC}$ undergo a sudden swap—$E_n^\mathrm{AB}$ jumps from 0 to 1 and $E_n^\mathrm{AC}$ from 1 to 0, realizing entanglement switching. This indicates that, in an optical‑lattice realization of the GAA model \cite{PhysRevLett.129.103401, PhysRevLett.126.040603, gauthier2016direct}, remote entanglement transfer and switching can be achieved simply by tuning $\alpha$ and $\lambda$ via external driving fields.

\textit{\tcm{Conclusion}---}
To summarize, we have proposed an alternative way to manipulate long-range entanglement using the quasiperiodic chains with tunable mobility edges. The indirect interactions between localized spins exhibit three distinct behaviors: (1) exponentially attenuating, (2) power-law decaying, and (3) selective enhancement for certain widely separated spin pairs when both energy scales coincide. We have shown that the third type of long-range interaction is originated from the critical states, and hence their anomalous behavior could be analyzed in terms of the local surrounding patterns. Furthermore, the experimentally tunable mobility edge serves as a switch of long-range entanglement: as it crosses the mobility edges, entanglement can be either adiabatically transferred from one to another or abruptly turned on or off, offering both smooth transfer and sharp switching functionalities.

Our results propose a new perspective in long-range entanglement control via pattern similarity and mobility edge which naturally occur in the GAA model. Moreover, they are broadly applicable to various incommensurate systems with mobility edges, emphasizing the importance of realizing and controlling such features across diverse experimental platforms. Optical lattices and photonic structures, with their tunable parameters, offer promising avenues for dynamical control of long-range entanglement \cite{vardeny2013optics, ozawa2019topological, li2017mobility, verresen2021efficiently}. Extending these insights to multi-spin networks \cite{hogben2011multiple, neschen1991efficient}, exotic entangled phases \cite{higgins2007entanglement}, and programmable entanglement \cite{cialdi2011programmable} presents an exciting direction for future work.

\section*{Acknowledgments}
This work was supported by National Research Foundation Grant (2021R1A2C109306013) and Nano Material Technology Development Program through the National Research Foundation of Korea(NRF) funded by Ministry of Science and ICT (RS-2023-00281839).

\bibliography{reference}

\begin{thebibliography}{67}%
\makeatletter
\providecommand \@ifxundefined [1]{%
 \@ifx{#1\undefined}
}%
\providecommand \@ifnum [1]{%
 \ifnum #1\expandafter \@firstoftwo
 \else \expandafter \@secondoftwo
 \fi
}%
\providecommand \@ifx [1]{%
 \ifx #1\expandafter \@firstoftwo
 \else \expandafter \@secondoftwo
 \fi
}%
\providecommand \natexlab [1]{#1}%
\providecommand \enquote  [1]{``#1''}%
\providecommand \bibnamefont  [1]{#1}%
\providecommand \bibfnamefont [1]{#1}%
\providecommand \citenamefont [1]{#1}%
\providecommand \href@noop [0]{\@secondoftwo}%
\providecommand \href [0]{\begingroup \@sanitize@url \@href}%
\providecommand \@href[1]{\@@startlink{#1}\@@href}%
\providecommand \@@href[1]{\endgroup#1\@@endlink}%
\providecommand \@sanitize@url [0]{\catcode `\\12\catcode `\$12\catcode `\&12\catcode `\#12\catcode `\^12\catcode `\_12\catcode `\%12\relax}%
\providecommand \@@startlink[1]{}%
\providecommand \@@endlink[0]{}%
\providecommand \url  [0]{\begingroup\@sanitize@url \@url }%
\providecommand \@url [1]{\endgroup\@href {#1}{\urlprefix }}%
\providecommand \urlprefix  [0]{URL }%
\providecommand \Eprint [0]{\href }%
\providecommand \doibase [0]{https://doi.org/}%
\providecommand \selectlanguage [0]{\@gobble}%
\providecommand \bibinfo  [0]{\@secondoftwo}%
\providecommand \bibfield  [0]{\@secondoftwo}%
\providecommand \translation [1]{[#1]}%
\providecommand \BibitemOpen [0]{}%
\providecommand \bibitemStop [0]{}%
\providecommand \bibitemNoStop [0]{.\EOS\space}%
\providecommand \EOS [0]{\spacefactor3000\relax}%
\providecommand \BibitemShut  [1]{\csname bibitem#1\endcsname}%
\let\auto@bib@innerbib\@empty
\bibitem [{\citenamefont {Horodecki}\ \emph {et~al.}(2009)\citenamefont {Horodecki}, \citenamefont {Horodecki}, \citenamefont {Horodecki},\ and\ \citenamefont {Horodecki}}]{RevModPhys.81.865}%
  \BibitemOpen
  \bibfield  {author} {\bibinfo {author} {\bibfnamefont {R.}~\bibnamefont {Horodecki}}, \bibinfo {author} {\bibfnamefont {P.}~\bibnamefont {Horodecki}}, \bibinfo {author} {\bibfnamefont {M.}~\bibnamefont {Horodecki}},\ and\ \bibinfo {author} {\bibfnamefont {K.}~\bibnamefont {Horodecki}},\ }\bibfield  {title} {\bibinfo {title} {Quantum entanglement},\ }\href@noop {} {\bibfield  {journal} {\bibinfo  {journal} {Rev. Mod. Phys.}\ }\textbf {\bibinfo {volume} {81}},\ \bibinfo {pages} {865} (\bibinfo {year} {2009})}\BibitemShut {NoStop}%
\bibitem [{\citenamefont {Audretsch}(2007)}]{audretsch2007entangled}%
  \BibitemOpen
  \bibfield  {author} {\bibinfo {author} {\bibfnamefont {J.}~\bibnamefont {Audretsch}},\ }\href@noop {} {\emph {\bibinfo {title} {Entangled systems: new directions in quantum physics}}}\ (\bibinfo  {publisher} {John Wiley \& Sons},\ \bibinfo {year} {2007})\BibitemShut {NoStop}%
\bibitem [{\citenamefont {Aharonov}(2000)}]{PhysRevA.62.062311}%
  \BibitemOpen
  \bibfield  {author} {\bibinfo {author} {\bibfnamefont {D.}~\bibnamefont {Aharonov}},\ }\bibfield  {title} {\bibinfo {title} {Quantum to classical phase transition in noisy quantum computers},\ }\href@noop {} {\bibfield  {journal} {\bibinfo  {journal} {Phys. Rev. A}\ }\textbf {\bibinfo {volume} {62}},\ \bibinfo {pages} {062311} (\bibinfo {year} {2000})}\BibitemShut {NoStop}%
\bibitem [{\citenamefont {Monroe}\ \emph {et~al.}(2014)\citenamefont {Monroe}, \citenamefont {Raussendorf}, \citenamefont {Ruthven}, \citenamefont {Brown}, \citenamefont {Maunz}, \citenamefont {Duan},\ and\ \citenamefont {Kim}}]{PhysRevA.89.022317}%
  \BibitemOpen
  \bibfield  {author} {\bibinfo {author} {\bibfnamefont {C.}~\bibnamefont {Monroe}}, \bibinfo {author} {\bibfnamefont {R.}~\bibnamefont {Raussendorf}}, \bibinfo {author} {\bibfnamefont {A.}~\bibnamefont {Ruthven}}, \bibinfo {author} {\bibfnamefont {K.~R.}\ \bibnamefont {Brown}}, \bibinfo {author} {\bibfnamefont {P.}~\bibnamefont {Maunz}}, \bibinfo {author} {\bibfnamefont {L.-M.}\ \bibnamefont {Duan}},\ and\ \bibinfo {author} {\bibfnamefont {J.}~\bibnamefont {Kim}},\ }\bibfield  {title} {\bibinfo {title} {Large-scale modular quantum-computer architecture with atomic memory and photonic interconnects},\ }\href@noop {} {\bibfield  {journal} {\bibinfo  {journal} {Phys. Rev. A}\ }\textbf {\bibinfo {volume} {89}},\ \bibinfo {pages} {022317} (\bibinfo {year} {2014})}\BibitemShut {NoStop}%
\bibitem [{\citenamefont {Long}\ and\ \citenamefont {Liu}(2002)}]{PhysRevA.65.032302}%
  \BibitemOpen
  \bibfield  {author} {\bibinfo {author} {\bibfnamefont {G.~L.}\ \bibnamefont {Long}}\ and\ \bibinfo {author} {\bibfnamefont {X.~S.}\ \bibnamefont {Liu}},\ }\bibfield  {title} {\bibinfo {title} {Theoretically efficient high-capacity quantum-key-distribution scheme},\ }\href@noop {} {\bibfield  {journal} {\bibinfo  {journal} {Phys. Rev. A}\ }\textbf {\bibinfo {volume} {65}},\ \bibinfo {pages} {032302} (\bibinfo {year} {2002})}\BibitemShut {NoStop}%
\bibitem [{\citenamefont {Kimble}(2008)}]{kimble2008quantum}%
  \BibitemOpen
  \bibfield  {author} {\bibinfo {author} {\bibfnamefont {H.~J.}\ \bibnamefont {Kimble}},\ }\bibfield  {title} {\bibinfo {title} {The quantum internet},\ }\href@noop {} {\bibfield  {journal} {\bibinfo  {journal} {Nature}\ }\textbf {\bibinfo {volume} {453}},\ \bibinfo {pages} {1023} (\bibinfo {year} {2008})}\BibitemShut {NoStop}%
\bibitem [{\citenamefont {Gisin}\ \emph {et~al.}(2002)\citenamefont {Gisin}, \citenamefont {Ribordy}, \citenamefont {Tittel},\ and\ \citenamefont {Zbinden}}]{RevModPhys.74.145}%
  \BibitemOpen
  \bibfield  {author} {\bibinfo {author} {\bibfnamefont {N.}~\bibnamefont {Gisin}}, \bibinfo {author} {\bibfnamefont {G.}~\bibnamefont {Ribordy}}, \bibinfo {author} {\bibfnamefont {W.}~\bibnamefont {Tittel}},\ and\ \bibinfo {author} {\bibfnamefont {H.}~\bibnamefont {Zbinden}},\ }\bibfield  {title} {\bibinfo {title} {Quantum cryptography},\ }\href@noop {} {\bibfield  {journal} {\bibinfo  {journal} {Rev. Mod. Phys.}\ }\textbf {\bibinfo {volume} {74}},\ \bibinfo {pages} {145} (\bibinfo {year} {2002})}\BibitemShut {NoStop}%
\bibitem [{\citenamefont {Bollinger}\ \emph {et~al.}(1996)\citenamefont {Bollinger}, \citenamefont {Itano}, \citenamefont {Wineland},\ and\ \citenamefont {Heinzen}}]{PhysRevA.54.R4649}%
  \BibitemOpen
  \bibfield  {author} {\bibinfo {author} {\bibfnamefont {J.~J.~.}\ \bibnamefont {Bollinger}}, \bibinfo {author} {\bibfnamefont {W.~M.}\ \bibnamefont {Itano}}, \bibinfo {author} {\bibfnamefont {D.~J.}\ \bibnamefont {Wineland}},\ and\ \bibinfo {author} {\bibfnamefont {D.~J.}\ \bibnamefont {Heinzen}},\ }\bibfield  {title} {\bibinfo {title} {Optimal frequency measurements with maximally correlated states},\ }\href@noop {} {\bibfield  {journal} {\bibinfo  {journal} {Phys. Rev. A}\ }\textbf {\bibinfo {volume} {54}},\ \bibinfo {pages} {R4649} (\bibinfo {year} {1996})}\BibitemShut {NoStop}%
\bibitem [{\citenamefont {Giovannetti}\ \emph {et~al.}(2006)\citenamefont {Giovannetti}, \citenamefont {Lloyd},\ and\ \citenamefont {Maccone}}]{PhysRevLett.96.010401}%
  \BibitemOpen
  \bibfield  {author} {\bibinfo {author} {\bibfnamefont {V.}~\bibnamefont {Giovannetti}}, \bibinfo {author} {\bibfnamefont {S.}~\bibnamefont {Lloyd}},\ and\ \bibinfo {author} {\bibfnamefont {L.}~\bibnamefont {Maccone}},\ }\bibfield  {title} {\bibinfo {title} {Quantum metrology},\ }\href@noop {} {\bibfield  {journal} {\bibinfo  {journal} {Phys. Rev. Lett.}\ }\textbf {\bibinfo {volume} {96}},\ \bibinfo {pages} {010401} (\bibinfo {year} {2006})}\BibitemShut {NoStop}%
\bibitem [{\citenamefont {Hughes}\ \emph {et~al.}(1996)\citenamefont {Hughes}, \citenamefont {James}, \citenamefont {Knill}, \citenamefont {Laflamme},\ and\ \citenamefont {Petschek}}]{PhysRevLett.77.3240}%
  \BibitemOpen
  \bibfield  {author} {\bibinfo {author} {\bibfnamefont {R.~J.}\ \bibnamefont {Hughes}}, \bibinfo {author} {\bibfnamefont {D.~F.~V.}\ \bibnamefont {James}}, \bibinfo {author} {\bibfnamefont {E.~H.}\ \bibnamefont {Knill}}, \bibinfo {author} {\bibfnamefont {R.}~\bibnamefont {Laflamme}},\ and\ \bibinfo {author} {\bibfnamefont {A.~G.}\ \bibnamefont {Petschek}},\ }\bibfield  {title} {\bibinfo {title} {Decoherence bounds on quantum computation with trapped ions},\ }\href@noop {} {\bibfield  {journal} {\bibinfo  {journal} {Phys. Rev. Lett.}\ }\textbf {\bibinfo {volume} {77}},\ \bibinfo {pages} {3240} (\bibinfo {year} {1996})}\BibitemShut {NoStop}%
\bibitem [{\citenamefont {Raimond}\ \emph {et~al.}(2001)\citenamefont {Raimond}, \citenamefont {Brune},\ and\ \citenamefont {Haroche}}]{RevModPhys.73.565}%
  \BibitemOpen
  \bibfield  {author} {\bibinfo {author} {\bibfnamefont {J.~M.}\ \bibnamefont {Raimond}}, \bibinfo {author} {\bibfnamefont {M.}~\bibnamefont {Brune}},\ and\ \bibinfo {author} {\bibfnamefont {S.}~\bibnamefont {Haroche}},\ }\bibfield  {title} {\bibinfo {title} {Manipulating quantum entanglement with atoms and photons in a cavity},\ }\href@noop {} {\bibfield  {journal} {\bibinfo  {journal} {Rev. Mod. Phys.}\ }\textbf {\bibinfo {volume} {73}},\ \bibinfo {pages} {565} (\bibinfo {year} {2001})}\BibitemShut {NoStop}%
\bibitem [{\citenamefont {Haroche}(1998)}]{haroche1998entanglement}%
  \BibitemOpen
  \bibfield  {author} {\bibinfo {author} {\bibfnamefont {S.}~\bibnamefont {Haroche}},\ }\bibfield  {title} {\bibinfo {title} {Entanglement, decoherence and the quantum/classical boundary},\ }\href@noop {} {\bibfield  {journal} {\bibinfo  {journal} {Physics today}\ }\textbf {\bibinfo {volume} {51}},\ \bibinfo {pages} {36} (\bibinfo {year} {1998})}\BibitemShut {NoStop}%
\bibitem [{\citenamefont {Tantivasadakarn}\ \emph {et~al.}(2024)\citenamefont {Tantivasadakarn}, \citenamefont {Thorngren}, \citenamefont {Vishwanath},\ and\ \citenamefont {Verresen}}]{tantivasadakarn2024long}%
  \BibitemOpen
  \bibfield  {author} {\bibinfo {author} {\bibfnamefont {N.}~\bibnamefont {Tantivasadakarn}}, \bibinfo {author} {\bibfnamefont {R.}~\bibnamefont {Thorngren}}, \bibinfo {author} {\bibfnamefont {A.}~\bibnamefont {Vishwanath}},\ and\ \bibinfo {author} {\bibfnamefont {R.}~\bibnamefont {Verresen}},\ }\bibfield  {title} {\bibinfo {title} {Long-range entanglement from measuring symmetry-protected topological phases},\ }\href@noop {} {\bibfield  {journal} {\bibinfo  {journal} {Physical Review X}\ }\textbf {\bibinfo {volume} {14}},\ \bibinfo {pages} {021040} (\bibinfo {year} {2024})}\BibitemShut {NoStop}%
\bibitem [{\citenamefont {Het{\'e}nyi}\ \emph {et~al.}(2022)\citenamefont {Het{\'e}nyi}, \citenamefont {Mook}, \citenamefont {Klinovaja},\ and\ \citenamefont {Loss}}]{hetenyi2022long}%
  \BibitemOpen
  \bibfield  {author} {\bibinfo {author} {\bibfnamefont {B.}~\bibnamefont {Het{\'e}nyi}}, \bibinfo {author} {\bibfnamefont {A.}~\bibnamefont {Mook}}, \bibinfo {author} {\bibfnamefont {J.}~\bibnamefont {Klinovaja}},\ and\ \bibinfo {author} {\bibfnamefont {D.}~\bibnamefont {Loss}},\ }\bibfield  {title} {\bibinfo {title} {Long-distance coupling of spin qubits via topological magnons},\ }\href@noop {} {\bibfield  {journal} {\bibinfo  {journal} {Physical Review B}\ }\textbf {\bibinfo {volume} {106}},\ \bibinfo {pages} {235409} (\bibinfo {year} {2022})}\BibitemShut {NoStop}%
\bibitem [{\citenamefont {Tang}\ \emph {et~al.}(2024)\citenamefont {Tang}, \citenamefont {Chen}, \citenamefont {Tang},\ and\ \citenamefont {Zhang}}]{tang2024topologically}%
  \BibitemOpen
  \bibfield  {author} {\bibinfo {author} {\bibfnamefont {Z.}~\bibnamefont {Tang}}, \bibinfo {author} {\bibfnamefont {T.}~\bibnamefont {Chen}}, \bibinfo {author} {\bibfnamefont {X.}~\bibnamefont {Tang}},\ and\ \bibinfo {author} {\bibfnamefont {X.}~\bibnamefont {Zhang}},\ }\bibfield  {title} {\bibinfo {title} {Topologically protected entanglement switching around exceptional points},\ }\href@noop {} {\bibfield  {journal} {\bibinfo  {journal} {Light: Science \& Applications}\ }\textbf {\bibinfo {volume} {13}},\ \bibinfo {pages} {167} (\bibinfo {year} {2024})}\BibitemShut {NoStop}%
\bibitem [{\citenamefont {Wang}\ \emph {et~al.}(2020)\citenamefont {Wang}, \citenamefont {Ren}, \citenamefont {Zhang}, \citenamefont {He},\ and\ \citenamefont {Zhang}}]{wang2020topologically}%
  \BibitemOpen
  \bibfield  {author} {\bibinfo {author} {\bibfnamefont {Y.}~\bibnamefont {Wang}}, \bibinfo {author} {\bibfnamefont {J.}~\bibnamefont {Ren}}, \bibinfo {author} {\bibfnamefont {W.}~\bibnamefont {Zhang}}, \bibinfo {author} {\bibfnamefont {L.}~\bibnamefont {He}},\ and\ \bibinfo {author} {\bibfnamefont {X.}~\bibnamefont {Zhang}},\ }\bibfield  {title} {\bibinfo {title} {Topologically protected strong coupling and entanglement between distant quantum emitters},\ }\href@noop {} {\bibfield  {journal} {\bibinfo  {journal} {Physical Review Applied}\ }\textbf {\bibinfo {volume} {14}},\ \bibinfo {pages} {054007} (\bibinfo {year} {2020})}\BibitemShut {NoStop}%
\bibitem [{\citenamefont {Mittal}\ \emph {et~al.}(2018)\citenamefont {Mittal}, \citenamefont {Goldschmidt},\ and\ \citenamefont {Hafezi}}]{mittal2018topological}%
  \BibitemOpen
  \bibfield  {author} {\bibinfo {author} {\bibfnamefont {S.}~\bibnamefont {Mittal}}, \bibinfo {author} {\bibfnamefont {E.~A.}\ \bibnamefont {Goldschmidt}},\ and\ \bibinfo {author} {\bibfnamefont {M.}~\bibnamefont {Hafezi}},\ }\bibfield  {title} {\bibinfo {title} {A topological source of quantum light},\ }\href@noop {} {\bibfield  {journal} {\bibinfo  {journal} {Nature}\ }\textbf {\bibinfo {volume} {561}},\ \bibinfo {pages} {502} (\bibinfo {year} {2018})}\BibitemShut {NoStop}%
\bibitem [{\citenamefont {Omran}\ \emph {et~al.}(2019)\citenamefont {Omran}, \citenamefont {Levine}, \citenamefont {Keesling}, \citenamefont {Semeghini}, \citenamefont {Wang}, \citenamefont {Ebadi}, \citenamefont {Bernien}, \citenamefont {Zibrov}, \citenamefont {Pichler}, \citenamefont {Choi} \emph {et~al.}}]{omran2019generation}%
  \BibitemOpen
  \bibfield  {author} {\bibinfo {author} {\bibfnamefont {A.}~\bibnamefont {Omran}}, \bibinfo {author} {\bibfnamefont {H.}~\bibnamefont {Levine}}, \bibinfo {author} {\bibfnamefont {A.}~\bibnamefont {Keesling}}, \bibinfo {author} {\bibfnamefont {G.}~\bibnamefont {Semeghini}}, \bibinfo {author} {\bibfnamefont {T.~T.}\ \bibnamefont {Wang}}, \bibinfo {author} {\bibfnamefont {S.}~\bibnamefont {Ebadi}}, \bibinfo {author} {\bibfnamefont {H.}~\bibnamefont {Bernien}}, \bibinfo {author} {\bibfnamefont {A.~S.}\ \bibnamefont {Zibrov}}, \bibinfo {author} {\bibfnamefont {H.}~\bibnamefont {Pichler}}, \bibinfo {author} {\bibfnamefont {S.}~\bibnamefont {Choi}}, \emph {et~al.},\ }\bibfield  {title} {\bibinfo {title} {Generation and manipulation of schr{\"o}dinger cat states in rydberg atom arrays},\ }\href@noop {} {\bibfield  {journal} {\bibinfo  {journal} {Science}\ }\textbf {\bibinfo {volume} {365}},\ \bibinfo {pages} {570} (\bibinfo {year} {2019})}\BibitemShut {NoStop}%
\bibitem [{\citenamefont {{\DH}or{\dj}evi{\'c}}\ \emph {et~al.}(2021)\citenamefont {{\DH}or{\dj}evi{\'c}}, \citenamefont {Samutpraphoot}, \citenamefont {Ocola}, \citenamefont {Bernien}, \citenamefont {Grinkemeyer}, \citenamefont {Dimitrova}, \citenamefont {Vuleti{\'c}},\ and\ \citenamefont {Lukin}}]{dhordjevic2021entanglement}%
  \BibitemOpen
  \bibfield  {author} {\bibinfo {author} {\bibfnamefont {T.}~\bibnamefont {{\DH}or{\dj}evi{\'c}}}, \bibinfo {author} {\bibfnamefont {P.}~\bibnamefont {Samutpraphoot}}, \bibinfo {author} {\bibfnamefont {P.~L.}\ \bibnamefont {Ocola}}, \bibinfo {author} {\bibfnamefont {H.}~\bibnamefont {Bernien}}, \bibinfo {author} {\bibfnamefont {B.}~\bibnamefont {Grinkemeyer}}, \bibinfo {author} {\bibfnamefont {I.}~\bibnamefont {Dimitrova}}, \bibinfo {author} {\bibfnamefont {V.}~\bibnamefont {Vuleti{\'c}}},\ and\ \bibinfo {author} {\bibfnamefont {M.~D.}\ \bibnamefont {Lukin}},\ }\bibfield  {title} {\bibinfo {title} {Entanglement transport and a nanophotonic interface for atoms in optical tweezers},\ }\href@noop {} {\bibfield  {journal} {\bibinfo  {journal} {Science}\ }\textbf {\bibinfo {volume} {373}},\ \bibinfo {pages} {1511} (\bibinfo {year} {2021})}\BibitemShut {NoStop}%
\bibitem [{\citenamefont {Shaw}\ \emph {et~al.}(2025)\citenamefont {Shaw}, \citenamefont {Scholl}, \citenamefont {Finkelstein}, \citenamefont {Tsai}, \citenamefont {Choi},\ and\ \citenamefont {Endres}}]{shaw2025erasure}%
  \BibitemOpen
  \bibfield  {author} {\bibinfo {author} {\bibfnamefont {A.~L.}\ \bibnamefont {Shaw}}, \bibinfo {author} {\bibfnamefont {P.}~\bibnamefont {Scholl}}, \bibinfo {author} {\bibfnamefont {R.}~\bibnamefont {Finkelstein}}, \bibinfo {author} {\bibfnamefont {R.~B.-S.}\ \bibnamefont {Tsai}}, \bibinfo {author} {\bibfnamefont {J.}~\bibnamefont {Choi}},\ and\ \bibinfo {author} {\bibfnamefont {M.}~\bibnamefont {Endres}},\ }\bibfield  {title} {\bibinfo {title} {Erasure cooling, control, and hyperentanglement of motion in optical tweezers},\ }\href@noop {} {\bibfield  {journal} {\bibinfo  {journal} {Science}\ }\textbf {\bibinfo {volume} {388}},\ \bibinfo {pages} {845} (\bibinfo {year} {2025})}\BibitemShut {NoStop}%
\bibitem [{\citenamefont {Hartung}\ \emph {et~al.}(2024)\citenamefont {Hartung}, \citenamefont {Seubert}, \citenamefont {Welte}, \citenamefont {Distante},\ and\ \citenamefont {Rempe}}]{hartung2024quantum}%
  \BibitemOpen
  \bibfield  {author} {\bibinfo {author} {\bibfnamefont {L.}~\bibnamefont {Hartung}}, \bibinfo {author} {\bibfnamefont {M.}~\bibnamefont {Seubert}}, \bibinfo {author} {\bibfnamefont {S.}~\bibnamefont {Welte}}, \bibinfo {author} {\bibfnamefont {E.}~\bibnamefont {Distante}},\ and\ \bibinfo {author} {\bibfnamefont {G.}~\bibnamefont {Rempe}},\ }\bibfield  {title} {\bibinfo {title} {A quantum-network register assembled with optical tweezers in an optical cavity},\ }\href@noop {} {\bibfield  {journal} {\bibinfo  {journal} {Science}\ }\textbf {\bibinfo {volume} {385}},\ \bibinfo {pages} {179} (\bibinfo {year} {2024})}\BibitemShut {NoStop}%
\bibitem [{\citenamefont {Iqbal}\ \emph {et~al.}(2024)\citenamefont {Iqbal}, \citenamefont {Tantivasadakarn}, \citenamefont {Gatterman}, \citenamefont {Gerber}, \citenamefont {Gilmore}, \citenamefont {Gresh}, \citenamefont {Hankin}, \citenamefont {Hewitt}, \citenamefont {Horst}, \citenamefont {Matheny} \emph {et~al.}}]{iqbal2024topological}%
  \BibitemOpen
  \bibfield  {author} {\bibinfo {author} {\bibfnamefont {M.}~\bibnamefont {Iqbal}}, \bibinfo {author} {\bibfnamefont {N.}~\bibnamefont {Tantivasadakarn}}, \bibinfo {author} {\bibfnamefont {T.~M.}\ \bibnamefont {Gatterman}}, \bibinfo {author} {\bibfnamefont {J.~A.}\ \bibnamefont {Gerber}}, \bibinfo {author} {\bibfnamefont {K.}~\bibnamefont {Gilmore}}, \bibinfo {author} {\bibfnamefont {D.}~\bibnamefont {Gresh}}, \bibinfo {author} {\bibfnamefont {A.}~\bibnamefont {Hankin}}, \bibinfo {author} {\bibfnamefont {N.}~\bibnamefont {Hewitt}}, \bibinfo {author} {\bibfnamefont {C.~V.}\ \bibnamefont {Horst}}, \bibinfo {author} {\bibfnamefont {M.}~\bibnamefont {Matheny}}, \emph {et~al.},\ }\bibfield  {title} {\bibinfo {title} {Topological order from measurements and feed-forward on a trapped ion quantum computer},\ }\href@noop {} {\bibfield  {journal} {\bibinfo  {journal} {Communications Physics}\ }\textbf {\bibinfo {volume} {7}},\ \bibinfo {pages} {205} (\bibinfo {year} {2024})}\BibitemShut {NoStop}%
\bibitem [{\citenamefont {B{\"a}umer}\ and\ \citenamefont {Woerner}(2025)}]{baumer2025measurement}%
  \BibitemOpen
  \bibfield  {author} {\bibinfo {author} {\bibfnamefont {E.}~\bibnamefont {B{\"a}umer}}\ and\ \bibinfo {author} {\bibfnamefont {S.}~\bibnamefont {Woerner}},\ }\bibfield  {title} {\bibinfo {title} {Measurement-based long-range entangling gates in constant depth},\ }\href@noop {} {\bibfield  {journal} {\bibinfo  {journal} {Physical Review Research}\ }\textbf {\bibinfo {volume} {7}},\ \bibinfo {pages} {023120} (\bibinfo {year} {2025})}\BibitemShut {NoStop}%
\bibitem [{\citenamefont {B{\"a}umer}\ \emph {et~al.}(2024)\citenamefont {B{\"a}umer}, \citenamefont {Tripathi}, \citenamefont {Wang}, \citenamefont {Rall}, \citenamefont {Chen}, \citenamefont {Majumder}, \citenamefont {Seif},\ and\ \citenamefont {Minev}}]{baumer2024efficient}%
  \BibitemOpen
  \bibfield  {author} {\bibinfo {author} {\bibfnamefont {E.}~\bibnamefont {B{\"a}umer}}, \bibinfo {author} {\bibfnamefont {V.}~\bibnamefont {Tripathi}}, \bibinfo {author} {\bibfnamefont {D.~S.}\ \bibnamefont {Wang}}, \bibinfo {author} {\bibfnamefont {P.}~\bibnamefont {Rall}}, \bibinfo {author} {\bibfnamefont {E.~H.}\ \bibnamefont {Chen}}, \bibinfo {author} {\bibfnamefont {S.}~\bibnamefont {Majumder}}, \bibinfo {author} {\bibfnamefont {A.}~\bibnamefont {Seif}},\ and\ \bibinfo {author} {\bibfnamefont {Z.~K.}\ \bibnamefont {Minev}},\ }\bibfield  {title} {\bibinfo {title} {Efficient long-range entanglement using dynamic circuits},\ }\href@noop {} {\bibfield  {journal} {\bibinfo  {journal} {PRX Quantum}\ }\textbf {\bibinfo {volume} {5}},\ \bibinfo {pages} {030339} (\bibinfo {year} {2024})}\BibitemShut {NoStop}%
\bibitem [{\citenamefont {Lu}\ \emph {et~al.}(2022)\citenamefont {Lu}, \citenamefont {Lessa}, \citenamefont {Kim},\ and\ \citenamefont {Hsieh}}]{lu2022measurement}%
  \BibitemOpen
  \bibfield  {author} {\bibinfo {author} {\bibfnamefont {T.-C.}\ \bibnamefont {Lu}}, \bibinfo {author} {\bibfnamefont {L.~A.}\ \bibnamefont {Lessa}}, \bibinfo {author} {\bibfnamefont {I.~H.}\ \bibnamefont {Kim}},\ and\ \bibinfo {author} {\bibfnamefont {T.~H.}\ \bibnamefont {Hsieh}},\ }\bibfield  {title} {\bibinfo {title} {Measurement as a shortcut to long-range entangled quantum matter},\ }\href@noop {} {\bibfield  {journal} {\bibinfo  {journal} {PRX Quantum}\ }\textbf {\bibinfo {volume} {3}},\ \bibinfo {pages} {040337} (\bibinfo {year} {2022})}\BibitemShut {NoStop}%
\bibitem [{\citenamefont {Sangouard}\ \emph {et~al.}(2011)\citenamefont {Sangouard}, \citenamefont {Simon}, \citenamefont {De~Riedmatten},\ and\ \citenamefont {Gisin}}]{sangouard2011quantum}%
  \BibitemOpen
  \bibfield  {author} {\bibinfo {author} {\bibfnamefont {N.}~\bibnamefont {Sangouard}}, \bibinfo {author} {\bibfnamefont {C.}~\bibnamefont {Simon}}, \bibinfo {author} {\bibfnamefont {H.}~\bibnamefont {De~Riedmatten}},\ and\ \bibinfo {author} {\bibfnamefont {N.}~\bibnamefont {Gisin}},\ }\bibfield  {title} {\bibinfo {title} {Quantum repeaters based on atomic ensembles and linear optics},\ }\href@noop {} {\bibfield  {journal} {\bibinfo  {journal} {Reviews of Modern Physics}\ }\textbf {\bibinfo {volume} {83}},\ \bibinfo {pages} {33} (\bibinfo {year} {2011})}\BibitemShut {NoStop}%
\bibitem [{\citenamefont {Agust\'{\i}}\ \emph {et~al.}(2022)\citenamefont {Agust\'{\i}}, \citenamefont {Minoguchi}, \citenamefont {Fink},\ and\ \citenamefont {Rabl}}]{PhysRevA.105.062454}%
  \BibitemOpen
  \bibfield  {author} {\bibinfo {author} {\bibfnamefont {J.}~\bibnamefont {Agust\'{\i}}}, \bibinfo {author} {\bibfnamefont {Y.}~\bibnamefont {Minoguchi}}, \bibinfo {author} {\bibfnamefont {J.~M.}\ \bibnamefont {Fink}},\ and\ \bibinfo {author} {\bibfnamefont {P.}~\bibnamefont {Rabl}},\ }\bibfield  {title} {\bibinfo {title} {Long-distance distribution of qubit-qubit entanglement using gaussian-correlated photonic beams},\ }\href@noop {} {\bibfield  {journal} {\bibinfo  {journal} {Phys. Rev. A}\ }\textbf {\bibinfo {volume} {105}},\ \bibinfo {pages} {062454} (\bibinfo {year} {2022})}\BibitemShut {NoStop}%
\bibitem [{\citenamefont {Krutyanskiy}\ \emph {et~al.}(2023)\citenamefont {Krutyanskiy}, \citenamefont {Galli}, \citenamefont {Krcmarsky}, \citenamefont {Baier}, \citenamefont {Fioretto}, \citenamefont {Pu}, \citenamefont {Mazloom}, \citenamefont {Sekatski}, \citenamefont {Canteri}, \citenamefont {Teller} \emph {et~al.}}]{krutyanskiy2023entanglement}%
  \BibitemOpen
  \bibfield  {author} {\bibinfo {author} {\bibfnamefont {V.}~\bibnamefont {Krutyanskiy}}, \bibinfo {author} {\bibfnamefont {M.}~\bibnamefont {Galli}}, \bibinfo {author} {\bibfnamefont {V.}~\bibnamefont {Krcmarsky}}, \bibinfo {author} {\bibfnamefont {S.}~\bibnamefont {Baier}}, \bibinfo {author} {\bibfnamefont {D.}~\bibnamefont {Fioretto}}, \bibinfo {author} {\bibfnamefont {Y.}~\bibnamefont {Pu}}, \bibinfo {author} {\bibfnamefont {A.}~\bibnamefont {Mazloom}}, \bibinfo {author} {\bibfnamefont {P.}~\bibnamefont {Sekatski}}, \bibinfo {author} {\bibfnamefont {M.}~\bibnamefont {Canteri}}, \bibinfo {author} {\bibfnamefont {M.}~\bibnamefont {Teller}}, \emph {et~al.},\ }\bibfield  {title} {\bibinfo {title} {Entanglement of trapped-ion qubits separated by 230 meters},\ }\href@noop {} {\bibfield  {journal} {\bibinfo  {journal} {Physical Review Letters}\ }\textbf {\bibinfo {volume} {130}},\ \bibinfo {pages} {050803} (\bibinfo {year} {2023})}\BibitemShut {NoStop}%
\bibitem [{\citenamefont {Qiu}\ \emph {et~al.}(2024)\citenamefont {Qiu}, \citenamefont {You}, \citenamefont {Guan}, \citenamefont {Du}, \citenamefont {He},\ and\ \citenamefont {Yang}}]{qiu2024efficient}%
  \BibitemOpen
  \bibfield  {author} {\bibinfo {author} {\bibfnamefont {H.}~\bibnamefont {Qiu}}, \bibinfo {author} {\bibfnamefont {Y.}~\bibnamefont {You}}, \bibinfo {author} {\bibfnamefont {X.-H.}\ \bibnamefont {Guan}}, \bibinfo {author} {\bibfnamefont {X.-J.}\ \bibnamefont {Du}}, \bibinfo {author} {\bibfnamefont {J.}~\bibnamefont {He}},\ and\ \bibinfo {author} {\bibfnamefont {Z.-J.}\ \bibnamefont {Yang}},\ }\bibfield  {title} {\bibinfo {title} {Efficient entanglement between two long-distance quantum emitters mediated by dark-gap-plasmon waveguides},\ }\href@noop {} {\bibfield  {journal} {\bibinfo  {journal} {Physical Review B}\ }\textbf {\bibinfo {volume} {110}},\ \bibinfo {pages} {075432} (\bibinfo {year} {2024})}\BibitemShut {NoStop}%
\bibitem [{\citenamefont {Burkard}\ \emph {et~al.}(2023)\citenamefont {Burkard}, \citenamefont {Ladd}, \citenamefont {Pan}, \citenamefont {Nichol},\ and\ \citenamefont {Petta}}]{burkard2023semiconductor}%
  \BibitemOpen
  \bibfield  {author} {\bibinfo {author} {\bibfnamefont {G.}~\bibnamefont {Burkard}}, \bibinfo {author} {\bibfnamefont {T.~D.}\ \bibnamefont {Ladd}}, \bibinfo {author} {\bibfnamefont {A.}~\bibnamefont {Pan}}, \bibinfo {author} {\bibfnamefont {J.~M.}\ \bibnamefont {Nichol}},\ and\ \bibinfo {author} {\bibfnamefont {J.~R.}\ \bibnamefont {Petta}},\ }\bibfield  {title} {\bibinfo {title} {Semiconductor spin qubits},\ }\href@noop {} {\bibfield  {journal} {\bibinfo  {journal} {Reviews of Modern Physics}\ }\textbf {\bibinfo {volume} {95}},\ \bibinfo {pages} {025003} (\bibinfo {year} {2023})}\BibitemShut {NoStop}%
\bibitem [{\citenamefont {Lei}\ \emph {et~al.}(2023)\citenamefont {Lei}, \citenamefont {Kimiaee~Asadi}, \citenamefont {Zhong}, \citenamefont {Kuzmich}, \citenamefont {Simon},\ and\ \citenamefont {Hosseini}}]{lei2023quantum}%
  \BibitemOpen
  \bibfield  {author} {\bibinfo {author} {\bibfnamefont {Y.}~\bibnamefont {Lei}}, \bibinfo {author} {\bibfnamefont {F.}~\bibnamefont {Kimiaee~Asadi}}, \bibinfo {author} {\bibfnamefont {T.}~\bibnamefont {Zhong}}, \bibinfo {author} {\bibfnamefont {A.}~\bibnamefont {Kuzmich}}, \bibinfo {author} {\bibfnamefont {C.}~\bibnamefont {Simon}},\ and\ \bibinfo {author} {\bibfnamefont {M.}~\bibnamefont {Hosseini}},\ }\bibfield  {title} {\bibinfo {title} {Quantum optical memory for entanglement distribution},\ }\href@noop {} {\bibfield  {journal} {\bibinfo  {journal} {Optica}\ }\textbf {\bibinfo {volume} {10}},\ \bibinfo {pages} {1511} (\bibinfo {year} {2023})}\BibitemShut {NoStop}%
\bibitem [{\citenamefont {Wei}\ \emph {et~al.}(2022)\citenamefont {Wei}, \citenamefont {Jing}, \citenamefont {Zhang}, \citenamefont {Liao}, \citenamefont {Yuan}, \citenamefont {Fan}, \citenamefont {Lyu}, \citenamefont {Zhou}, \citenamefont {Wang}, \citenamefont {Deng} \emph {et~al.}}]{wei2022towards}%
  \BibitemOpen
  \bibfield  {author} {\bibinfo {author} {\bibfnamefont {S.-H.}\ \bibnamefont {Wei}}, \bibinfo {author} {\bibfnamefont {B.}~\bibnamefont {Jing}}, \bibinfo {author} {\bibfnamefont {X.-Y.}\ \bibnamefont {Zhang}}, \bibinfo {author} {\bibfnamefont {J.-Y.}\ \bibnamefont {Liao}}, \bibinfo {author} {\bibfnamefont {C.-Z.}\ \bibnamefont {Yuan}}, \bibinfo {author} {\bibfnamefont {B.-Y.}\ \bibnamefont {Fan}}, \bibinfo {author} {\bibfnamefont {C.}~\bibnamefont {Lyu}}, \bibinfo {author} {\bibfnamefont {D.-L.}\ \bibnamefont {Zhou}}, \bibinfo {author} {\bibfnamefont {Y.}~\bibnamefont {Wang}}, \bibinfo {author} {\bibfnamefont {G.-W.}\ \bibnamefont {Deng}}, \emph {et~al.},\ }\bibfield  {title} {\bibinfo {title} {Towards real-world quantum networks: a review},\ }\href@noop {} {\bibfield  {journal} {\bibinfo  {journal} {Laser \& Photonics Reviews}\ }\textbf {\bibinfo {volume} {16}},\ \bibinfo {pages} {2100219} (\bibinfo {year} {2022})}\BibitemShut {NoStop}%
\bibitem [{\citenamefont {Hu}\ \emph {et~al.}(2021)\citenamefont {Hu}, \citenamefont {Huang}, \citenamefont {Sheng}, \citenamefont {Zhou}, \citenamefont {Liu}, \citenamefont {Guo}, \citenamefont {Zhang}, \citenamefont {Xing}, \citenamefont {Huang}, \citenamefont {Li},\ and\ \citenamefont {Guo}}]{PhysRevLett.126.010503}%
  \BibitemOpen
  \bibfield  {author} {\bibinfo {author} {\bibfnamefont {X.-M.}\ \bibnamefont {Hu}}, \bibinfo {author} {\bibfnamefont {C.-X.}\ \bibnamefont {Huang}}, \bibinfo {author} {\bibfnamefont {Y.-B.}\ \bibnamefont {Sheng}}, \bibinfo {author} {\bibfnamefont {L.}~\bibnamefont {Zhou}}, \bibinfo {author} {\bibfnamefont {B.-H.}\ \bibnamefont {Liu}}, \bibinfo {author} {\bibfnamefont {Y.}~\bibnamefont {Guo}}, \bibinfo {author} {\bibfnamefont {C.}~\bibnamefont {Zhang}}, \bibinfo {author} {\bibfnamefont {W.-B.}\ \bibnamefont {Xing}}, \bibinfo {author} {\bibfnamefont {Y.-F.}\ \bibnamefont {Huang}}, \bibinfo {author} {\bibfnamefont {C.-F.}\ \bibnamefont {Li}},\ and\ \bibinfo {author} {\bibfnamefont {G.-C.}\ \bibnamefont {Guo}},\ }\bibfield  {title} {\bibinfo {title} {Long-distance entanglement purification for quantum communication},\ }\href@noop {} {\bibfield  {journal} {\bibinfo  {journal} {Phys. Rev. Lett.}\ }\textbf {\bibinfo {volume} {126}},\ \bibinfo {pages} {010503} (\bibinfo {year} {2021})}\BibitemShut {NoStop}%
\bibitem [{\citenamefont {Ursin}\ \emph {et~al.}(2007)\citenamefont {Ursin}, \citenamefont {Tiefenbacher}, \citenamefont {Schmitt-Manderbach}, \citenamefont {Weier}, \citenamefont {Scheidl}, \citenamefont {Lindenthal}, \citenamefont {Blauensteiner}, \citenamefont {Jennewein}, \citenamefont {Perdigues}, \citenamefont {Trojek} \emph {et~al.}}]{ursin2007entanglement}%
  \BibitemOpen
  \bibfield  {author} {\bibinfo {author} {\bibfnamefont {R.}~\bibnamefont {Ursin}}, \bibinfo {author} {\bibfnamefont {F.}~\bibnamefont {Tiefenbacher}}, \bibinfo {author} {\bibfnamefont {T.}~\bibnamefont {Schmitt-Manderbach}}, \bibinfo {author} {\bibfnamefont {H.}~\bibnamefont {Weier}}, \bibinfo {author} {\bibfnamefont {T.}~\bibnamefont {Scheidl}}, \bibinfo {author} {\bibfnamefont {M.}~\bibnamefont {Lindenthal}}, \bibinfo {author} {\bibfnamefont {B.}~\bibnamefont {Blauensteiner}}, \bibinfo {author} {\bibfnamefont {T.}~\bibnamefont {Jennewein}}, \bibinfo {author} {\bibfnamefont {J.}~\bibnamefont {Perdigues}}, \bibinfo {author} {\bibfnamefont {P.}~\bibnamefont {Trojek}}, \emph {et~al.},\ }\bibfield  {title} {\bibinfo {title} {Entanglement-based quantum communication over 144 km},\ }\href@noop {} {\bibfield  {journal} {\bibinfo  {journal} {Nature physics}\ }\textbf {\bibinfo {volume} {3}},\ \bibinfo {pages} {481} (\bibinfo {year} {2007})}\BibitemShut {NoStop}%
\bibitem [{\citenamefont {Yuan}\ \emph {et~al.}(2010)\citenamefont {Yuan}, \citenamefont {Bao}, \citenamefont {Lu}, \citenamefont {Zhang}, \citenamefont {Peng},\ and\ \citenamefont {Pan}}]{yuan2010entangled}%
  \BibitemOpen
  \bibfield  {author} {\bibinfo {author} {\bibfnamefont {Z.-S.}\ \bibnamefont {Yuan}}, \bibinfo {author} {\bibfnamefont {X.-H.}\ \bibnamefont {Bao}}, \bibinfo {author} {\bibfnamefont {C.-Y.}\ \bibnamefont {Lu}}, \bibinfo {author} {\bibfnamefont {J.}~\bibnamefont {Zhang}}, \bibinfo {author} {\bibfnamefont {C.-Z.}\ \bibnamefont {Peng}},\ and\ \bibinfo {author} {\bibfnamefont {J.-W.}\ \bibnamefont {Pan}},\ }\bibfield  {title} {\bibinfo {title} {Entangled photons and quantum communication},\ }\href@noop {} {\bibfield  {journal} {\bibinfo  {journal} {Physics Reports}\ }\textbf {\bibinfo {volume} {497}},\ \bibinfo {pages} {1} (\bibinfo {year} {2010})}\BibitemShut {NoStop}%
\bibitem [{\citenamefont {Duan}\ \emph {et~al.}(2001)\citenamefont {Duan}, \citenamefont {Lukin}, \citenamefont {Cirac},\ and\ \citenamefont {Zoller}}]{duan2001long}%
  \BibitemOpen
  \bibfield  {author} {\bibinfo {author} {\bibfnamefont {L.-M.}\ \bibnamefont {Duan}}, \bibinfo {author} {\bibfnamefont {M.~D.}\ \bibnamefont {Lukin}}, \bibinfo {author} {\bibfnamefont {J.~I.}\ \bibnamefont {Cirac}},\ and\ \bibinfo {author} {\bibfnamefont {P.}~\bibnamefont {Zoller}},\ }\bibfield  {title} {\bibinfo {title} {Long-distance quantum communication with atomic ensembles and linear optics},\ }\href@noop {} {\bibfield  {journal} {\bibinfo  {journal} {Nature}\ }\textbf {\bibinfo {volume} {414}},\ \bibinfo {pages} {413} (\bibinfo {year} {2001})}\BibitemShut {NoStop}%
\bibitem [{\citenamefont {Jeon}\ and\ \citenamefont {Lee}(2025{\natexlab{a}})}]{PhysRevB.111.L020411}%
  \BibitemOpen
  \bibfield  {author} {\bibinfo {author} {\bibfnamefont {J.}~\bibnamefont {Jeon}}\ and\ \bibinfo {author} {\bibfnamefont {S.}~\bibnamefont {Lee}},\ }\bibfield  {title} {\bibinfo {title} {Hidden hyperspace geometry and long-distance quantum coupling},\ }\href@noop {} {\bibfield  {journal} {\bibinfo  {journal} {Phys. Rev. B}\ }\textbf {\bibinfo {volume} {111}},\ \bibinfo {pages} {L020411} (\bibinfo {year} {2025}{\natexlab{a}})}\BibitemShut {NoStop}%
\bibitem [{\citenamefont {Wilkinson}(1984)}]{wilkinson1984critical}%
  \BibitemOpen
  \bibfield  {author} {\bibinfo {author} {\bibfnamefont {M.}~\bibnamefont {Wilkinson}},\ }\bibfield  {title} {\bibinfo {title} {Critical properties of electron eigenstates in incommensurate systems},\ }\href@noop {} {\bibfield  {journal} {\bibinfo  {journal} {Proceedings of the Royal Society of London. A. Mathematical and Physical Sciences}\ }\textbf {\bibinfo {volume} {391}},\ \bibinfo {pages} {305} (\bibinfo {year} {1984})}\BibitemShut {NoStop}%
\bibitem [{\citenamefont {Kohmoto}(1983)}]{PhysRevLett.51.1198}%
  \BibitemOpen
  \bibfield  {author} {\bibinfo {author} {\bibfnamefont {M.}~\bibnamefont {Kohmoto}},\ }\bibfield  {title} {\bibinfo {title} {Metal-insulator transition and scaling for incommensurate systems},\ }\href@noop {} {\bibfield  {journal} {\bibinfo  {journal} {Phys. Rev. Lett.}\ }\textbf {\bibinfo {volume} {51}},\ \bibinfo {pages} {1198} (\bibinfo {year} {1983})}\BibitemShut {NoStop}%
\bibitem [{\citenamefont {Aubry}\ and\ \citenamefont {Andr{\'e}}()}]{aubry1980analyticity}%
  \BibitemOpen
  \bibfield  {author} {\bibinfo {author} {\bibfnamefont {S.}~\bibnamefont {Aubry}}\ and\ \bibinfo {author} {\bibfnamefont {G.}~\bibnamefont {Andr{\'e}}},\ }\bibfield  {title} {\bibinfo {title} {Analyticity breaking and anderson localization in incommensurate lattices},\ }\href@noop {} {\bibfield  {journal} {\bibinfo  {journal} {Ann. Israel Phys. Soc}\ }\textbf {\bibinfo {volume} {3}},\ \bibinfo {pages} {18}}\BibitemShut {NoStop}%
\bibitem [{\citenamefont {Shklovskii}\ and\ \citenamefont {Efros}(2013)}]{shklovskii2013electronic}%
  \BibitemOpen
  \bibfield  {author} {\bibinfo {author} {\bibfnamefont {B.~I.}\ \bibnamefont {Shklovskii}}\ and\ \bibinfo {author} {\bibfnamefont {A.~L.}\ \bibnamefont {Efros}},\ }\href@noop {} {\emph {\bibinfo {title} {Electronic properties of doped semiconductors}}},\ Vol.~\bibinfo {volume} {45}\ (\bibinfo  {publisher} {Springer Science \& Business Media},\ \bibinfo {year} {2013})\BibitemShut {NoStop}%
\bibitem [{\citenamefont {Ganeshan}\ \emph {et~al.}(2015)\citenamefont {Ganeshan}, \citenamefont {Pixley},\ and\ \citenamefont {Das~Sarma}}]{PhysRevLett.114.146601}%
  \BibitemOpen
  \bibfield  {author} {\bibinfo {author} {\bibfnamefont {S.}~\bibnamefont {Ganeshan}}, \bibinfo {author} {\bibfnamefont {J.~H.}\ \bibnamefont {Pixley}},\ and\ \bibinfo {author} {\bibfnamefont {S.}~\bibnamefont {Das~Sarma}},\ }\bibfield  {title} {\bibinfo {title} {Nearest neighbor tight binding models with an exact mobility edge in one dimension},\ }\href@noop {} {\bibfield  {journal} {\bibinfo  {journal} {Phys. Rev. Lett.}\ }\textbf {\bibinfo {volume} {114}},\ \bibinfo {pages} {146601} (\bibinfo {year} {2015})}\BibitemShut {NoStop}%
\bibitem [{\citenamefont {Biddle}\ and\ \citenamefont {Das~Sarma}(2010)}]{PhysRevLett.104.070601}%
  \BibitemOpen
  \bibfield  {author} {\bibinfo {author} {\bibfnamefont {J.}~\bibnamefont {Biddle}}\ and\ \bibinfo {author} {\bibfnamefont {S.}~\bibnamefont {Das~Sarma}},\ }\bibfield  {title} {\bibinfo {title} {Predicted mobility edges in one-dimensional incommensurate optical lattices: An exactly solvable model of anderson localization},\ }\href@noop {} {\bibfield  {journal} {\bibinfo  {journal} {Phys. Rev. Lett.}\ }\textbf {\bibinfo {volume} {104}},\ \bibinfo {pages} {070601} (\bibinfo {year} {2010})}\BibitemShut {NoStop}%
\bibitem [{\citenamefont {An}\ \emph {et~al.}(2021)\citenamefont {An}, \citenamefont {Padavi\ifmmode~\acute{c}\else \'{c}\fi{}}, \citenamefont {Meier}, \citenamefont {Hegde}, \citenamefont {Ganeshan}, \citenamefont {Pixley}, \citenamefont {Vishveshwara},\ and\ \citenamefont {Gadway}}]{PhysRevLett.126.040603}%
  \BibitemOpen
  \bibfield  {author} {\bibinfo {author} {\bibfnamefont {F.~A.}\ \bibnamefont {An}}, \bibinfo {author} {\bibfnamefont {K.}~\bibnamefont {Padavi\ifmmode~\acute{c}\else \'{c}\fi{}}}, \bibinfo {author} {\bibfnamefont {E.~J.}\ \bibnamefont {Meier}}, \bibinfo {author} {\bibfnamefont {S.}~\bibnamefont {Hegde}}, \bibinfo {author} {\bibfnamefont {S.}~\bibnamefont {Ganeshan}}, \bibinfo {author} {\bibfnamefont {J.~H.}\ \bibnamefont {Pixley}}, \bibinfo {author} {\bibfnamefont {S.}~\bibnamefont {Vishveshwara}},\ and\ \bibinfo {author} {\bibfnamefont {B.}~\bibnamefont {Gadway}},\ }\bibfield  {title} {\bibinfo {title} {Interactions and mobility edges: Observing the generalized aubry-andr\'e model},\ }\href@noop {} {\bibfield  {journal} {\bibinfo  {journal} {Phys. Rev. Lett.}\ }\textbf {\bibinfo {volume} {126}},\ \bibinfo {pages} {040603} (\bibinfo {year} {2021})}\BibitemShut {NoStop}%
\bibitem [{\citenamefont {Wang}\ \emph {et~al.}(2022)\citenamefont {Wang}, \citenamefont {Zhang}, \citenamefont {Li}, \citenamefont {Wu}, \citenamefont {Liu}, \citenamefont {Mei}, \citenamefont {Hu}, \citenamefont {Xiao}, \citenamefont {Ma}, \citenamefont {Chin},\ and\ \citenamefont {Jia}}]{PhysRevLett.129.103401}%
  \BibitemOpen
  \bibfield  {author} {\bibinfo {author} {\bibfnamefont {Y.}~\bibnamefont {Wang}}, \bibinfo {author} {\bibfnamefont {J.-H.}\ \bibnamefont {Zhang}}, \bibinfo {author} {\bibfnamefont {Y.}~\bibnamefont {Li}}, \bibinfo {author} {\bibfnamefont {J.}~\bibnamefont {Wu}}, \bibinfo {author} {\bibfnamefont {W.}~\bibnamefont {Liu}}, \bibinfo {author} {\bibfnamefont {F.}~\bibnamefont {Mei}}, \bibinfo {author} {\bibfnamefont {Y.}~\bibnamefont {Hu}}, \bibinfo {author} {\bibfnamefont {L.}~\bibnamefont {Xiao}}, \bibinfo {author} {\bibfnamefont {J.}~\bibnamefont {Ma}}, \bibinfo {author} {\bibfnamefont {C.}~\bibnamefont {Chin}},\ and\ \bibinfo {author} {\bibfnamefont {S.}~\bibnamefont {Jia}},\ }\bibfield  {title} {\bibinfo {title} {Observation of interaction-induced mobility edge in an atomic aubry-andr\'e wire},\ }\href@noop {} {\bibfield  {journal} {\bibinfo  {journal} {Phys. Rev. Lett.}\ }\textbf {\bibinfo {volume} {129}},\ \bibinfo {pages} {103401} (\bibinfo {year} {2022})}\BibitemShut {NoStop}%
\bibitem [{\citenamefont {Qiu}\ \emph {et~al.}(2020)\citenamefont {Qiu}, \citenamefont {Zou}, \citenamefont {Qi},\ and\ \citenamefont {Li}}]{qiu2020precise}%
  \BibitemOpen
  \bibfield  {author} {\bibinfo {author} {\bibfnamefont {X.}~\bibnamefont {Qiu}}, \bibinfo {author} {\bibfnamefont {J.}~\bibnamefont {Zou}}, \bibinfo {author} {\bibfnamefont {X.}~\bibnamefont {Qi}},\ and\ \bibinfo {author} {\bibfnamefont {X.}~\bibnamefont {Li}},\ }\bibfield  {title} {\bibinfo {title} {Precise programmable quantum simulations with optical lattices},\ }\href@noop {} {\bibfield  {journal} {\bibinfo  {journal} {npj Quantum Information}\ }\textbf {\bibinfo {volume} {6}},\ \bibinfo {pages} {87} (\bibinfo {year} {2020})}\BibitemShut {NoStop}%
\bibitem [{\citenamefont {Lukin}\ \emph {et~al.}(2019)\citenamefont {Lukin}, \citenamefont {Rispoli}, \citenamefont {Schittko}, \citenamefont {Tai}, \citenamefont {Kaufman}, \citenamefont {Choi}, \citenamefont {Khemani}, \citenamefont {L{\'e}onard},\ and\ \citenamefont {Greiner}}]{lukin2019probing}%
  \BibitemOpen
  \bibfield  {author} {\bibinfo {author} {\bibfnamefont {A.}~\bibnamefont {Lukin}}, \bibinfo {author} {\bibfnamefont {M.}~\bibnamefont {Rispoli}}, \bibinfo {author} {\bibfnamefont {R.}~\bibnamefont {Schittko}}, \bibinfo {author} {\bibfnamefont {M.~E.}\ \bibnamefont {Tai}}, \bibinfo {author} {\bibfnamefont {A.~M.}\ \bibnamefont {Kaufman}}, \bibinfo {author} {\bibfnamefont {S.}~\bibnamefont {Choi}}, \bibinfo {author} {\bibfnamefont {V.}~\bibnamefont {Khemani}}, \bibinfo {author} {\bibfnamefont {J.}~\bibnamefont {L{\'e}onard}},\ and\ \bibinfo {author} {\bibfnamefont {M.}~\bibnamefont {Greiner}},\ }\bibfield  {title} {\bibinfo {title} {Probing entanglement in a many-body--localized system},\ }\href@noop {} {\bibfield  {journal} {\bibinfo  {journal} {Science}\ }\textbf {\bibinfo {volume} {364}},\ \bibinfo {pages} {256} (\bibinfo {year} {2019})}\BibitemShut {NoStop}%
\bibitem [{\citenamefont {Gauthier}\ \emph {et~al.}(2016)\citenamefont {Gauthier}, \citenamefont {Lenton}, \citenamefont {McKay~Parry}, \citenamefont {Baker}, \citenamefont {Davis}, \citenamefont {Rubinsztein-Dunlop},\ and\ \citenamefont {Neely}}]{gauthier2016direct}%
  \BibitemOpen
  \bibfield  {author} {\bibinfo {author} {\bibfnamefont {G.}~\bibnamefont {Gauthier}}, \bibinfo {author} {\bibfnamefont {I.}~\bibnamefont {Lenton}}, \bibinfo {author} {\bibfnamefont {N.}~\bibnamefont {McKay~Parry}}, \bibinfo {author} {\bibfnamefont {M.}~\bibnamefont {Baker}}, \bibinfo {author} {\bibfnamefont {M.}~\bibnamefont {Davis}}, \bibinfo {author} {\bibfnamefont {H.}~\bibnamefont {Rubinsztein-Dunlop}},\ and\ \bibinfo {author} {\bibfnamefont {T.}~\bibnamefont {Neely}},\ }\bibfield  {title} {\bibinfo {title} {Direct imaging of a digital-micromirror device for configurable microscopic optical potentials},\ }\href@noop {} {\bibfield  {journal} {\bibinfo  {journal} {Optica}\ }\textbf {\bibinfo {volume} {3}},\ \bibinfo {pages} {1136} (\bibinfo {year} {2016})}\BibitemShut {NoStop}%
\bibitem [{\citenamefont {Ruderman}\ and\ \citenamefont {Kittel}(1954)}]{PhysRev.96.99}%
  \BibitemOpen
  \bibfield  {author} {\bibinfo {author} {\bibfnamefont {M.~A.}\ \bibnamefont {Ruderman}}\ and\ \bibinfo {author} {\bibfnamefont {C.}~\bibnamefont {Kittel}},\ }\bibfield  {title} {\bibinfo {title} {Indirect exchange coupling of nuclear magnetic moments by conduction electrons},\ }\href@noop {} {\bibfield  {journal} {\bibinfo  {journal} {Phys. Rev.}\ }\textbf {\bibinfo {volume} {96}},\ \bibinfo {pages} {99} (\bibinfo {year} {1954})}\BibitemShut {NoStop}%
\bibitem [{\citenamefont {Yafet}(1987)}]{PhysRevB.36.3948}%
  \BibitemOpen
  \bibfield  {author} {\bibinfo {author} {\bibfnamefont {Y.}~\bibnamefont {Yafet}},\ }\bibfield  {title} {\bibinfo {title} {Ruderman-kittel-kasuya-yosida range function of a one-dimensional free-electron gas},\ }\href@noop {} {\bibfield  {journal} {\bibinfo  {journal} {Phys. Rev. B}\ }\textbf {\bibinfo {volume} {36}},\ \bibinfo {pages} {3948} (\bibinfo {year} {1987})}\BibitemShut {NoStop}%
\bibitem [{\citenamefont {Bauer}\ \emph {et~al.}(1990)\citenamefont {Bauer}, \citenamefont {Chang},\ and\ \citenamefont {Skinner}}]{bauer1990correlation}%
  \BibitemOpen
  \bibfield  {author} {\bibinfo {author} {\bibfnamefont {J.}~\bibnamefont {Bauer}}, \bibinfo {author} {\bibfnamefont {T.-M.}\ \bibnamefont {Chang}},\ and\ \bibinfo {author} {\bibfnamefont {J.}~\bibnamefont {Skinner}},\ }\bibfield  {title} {\bibinfo {title} {Correlation length and inverse-participation-ratio exponents and multifractal structure for anderson localization},\ }\href@noop {} {\bibfield  {journal} {\bibinfo  {journal} {Physical Review B}\ }\textbf {\bibinfo {volume} {42}},\ \bibinfo {pages} {8121} (\bibinfo {year} {1990})}\BibitemShut {NoStop}%
\bibitem [{\citenamefont {Niimi}\ \emph {et~al.}(2006)\citenamefont {Niimi}, \citenamefont {Matsui}, \citenamefont {Kambara}, \citenamefont {Tagami}, \citenamefont {Tsukada},\ and\ \citenamefont {Fukuyama}}]{niimi2006scanning}%
  \BibitemOpen
  \bibfield  {author} {\bibinfo {author} {\bibfnamefont {Y.}~\bibnamefont {Niimi}}, \bibinfo {author} {\bibfnamefont {T.}~\bibnamefont {Matsui}}, \bibinfo {author} {\bibfnamefont {H.}~\bibnamefont {Kambara}}, \bibinfo {author} {\bibfnamefont {K.}~\bibnamefont {Tagami}}, \bibinfo {author} {\bibfnamefont {M.}~\bibnamefont {Tsukada}},\ and\ \bibinfo {author} {\bibfnamefont {H.}~\bibnamefont {Fukuyama}},\ }\bibfield  {title} {\bibinfo {title} {Scanning tunneling microscopy and spectroscopy of the electronic local density of states of graphite surfaces near monoatomic step edges},\ }\href@noop {} {\bibfield  {journal} {\bibinfo  {journal} {Physical Review B—Condensed Matter and Materials Physics}\ }\textbf {\bibinfo {volume} {73}},\ \bibinfo {pages} {085421} (\bibinfo {year} {2006})}\BibitemShut {NoStop}%
\bibitem [{\citenamefont {Chicanne}\ \emph {et~al.}(2002)\citenamefont {Chicanne}, \citenamefont {David}, \citenamefont {Quidant}, \citenamefont {Weeber}, \citenamefont {Lacroute}, \citenamefont {Bourillot}, \citenamefont {Dereux}, \citenamefont {Des~Francs},\ and\ \citenamefont {Girard}}]{chicanne2002imaging}%
  \BibitemOpen
  \bibfield  {author} {\bibinfo {author} {\bibfnamefont {C.}~\bibnamefont {Chicanne}}, \bibinfo {author} {\bibfnamefont {T.}~\bibnamefont {David}}, \bibinfo {author} {\bibfnamefont {R.}~\bibnamefont {Quidant}}, \bibinfo {author} {\bibfnamefont {J.-C.}\ \bibnamefont {Weeber}}, \bibinfo {author} {\bibfnamefont {Y.}~\bibnamefont {Lacroute}}, \bibinfo {author} {\bibfnamefont {E.}~\bibnamefont {Bourillot}}, \bibinfo {author} {\bibfnamefont {A.}~\bibnamefont {Dereux}}, \bibinfo {author} {\bibfnamefont {G.~C.}\ \bibnamefont {Des~Francs}},\ and\ \bibinfo {author} {\bibfnamefont {C.}~\bibnamefont {Girard}},\ }\bibfield  {title} {\bibinfo {title} {Imaging the local density of states of optical corrals},\ }\href@noop {} {\bibfield  {journal} {\bibinfo  {journal} {Physical review letters}\ }\textbf {\bibinfo {volume} {88}},\ \bibinfo {pages} {097402} (\bibinfo {year} {2002})}\BibitemShut {NoStop}%
\bibitem [{\citenamefont {Gisin}\ and\ \citenamefont {Thew}(2007)}]{gisin2007quantum}%
  \BibitemOpen
  \bibfield  {author} {\bibinfo {author} {\bibfnamefont {N.}~\bibnamefont {Gisin}}\ and\ \bibinfo {author} {\bibfnamefont {R.}~\bibnamefont {Thew}},\ }\bibfield  {title} {\bibinfo {title} {Quantum communication},\ }\href@noop {} {\bibfield  {journal} {\bibinfo  {journal} {Nature photonics}\ }\textbf {\bibinfo {volume} {1}},\ \bibinfo {pages} {165} (\bibinfo {year} {2007})}\BibitemShut {NoStop}%
\bibitem [{\citenamefont {Al-Qasimi}\ and\ \citenamefont {James}(2008)}]{PhysRevA.77.012117}%
  \BibitemOpen
  \bibfield  {author} {\bibinfo {author} {\bibfnamefont {A.}~\bibnamefont {Al-Qasimi}}\ and\ \bibinfo {author} {\bibfnamefont {D.~F.~V.}\ \bibnamefont {James}},\ }\bibfield  {title} {\bibinfo {title} {Sudden death of entanglement at finite temperature},\ }\href@noop {} {\bibfield  {journal} {\bibinfo  {journal} {Phys. Rev. A}\ }\textbf {\bibinfo {volume} {77}},\ \bibinfo {pages} {012117} (\bibinfo {year} {2008})}\BibitemShut {NoStop}%
\bibitem [{\citenamefont {Jeon}\ and\ \citenamefont {Lee}(2025{\natexlab{b}})}]{jeon2025immortal}%
  \BibitemOpen
  \bibfield  {author} {\bibinfo {author} {\bibfnamefont {J.}~\bibnamefont {Jeon}}\ and\ \bibinfo {author} {\bibfnamefont {S.}~\bibnamefont {Lee}},\ }\bibfield  {title} {\bibinfo {title} {Immortal quantum correlation in quasiperiodic quasi-one-dimensional systems},\ }\href@noop {} {\bibfield  {journal} {\bibinfo  {journal} {Physical Review B}\ }\textbf {\bibinfo {volume} {111}},\ \bibinfo {pages} {L161117} (\bibinfo {year} {2025}{\natexlab{b}})}\BibitemShut {NoStop}%
\bibitem [{\citenamefont {Vidal}\ and\ \citenamefont {Werner}(2002)}]{PhysRevA.65.032314}%
  \BibitemOpen
  \bibfield  {author} {\bibinfo {author} {\bibfnamefont {G.}~\bibnamefont {Vidal}}\ and\ \bibinfo {author} {\bibfnamefont {R.~F.}\ \bibnamefont {Werner}},\ }\bibfield  {title} {\bibinfo {title} {Computable measure of entanglement},\ }\href@noop {} {\bibfield  {journal} {\bibinfo  {journal} {Phys. Rev. A}\ }\textbf {\bibinfo {volume} {65}},\ \bibinfo {pages} {032314} (\bibinfo {year} {2002})}\BibitemShut {NoStop}%
\bibitem [{\citenamefont {Horodecki}(2001)}]{horodecki2001entanglement}%
  \BibitemOpen
  \bibfield  {author} {\bibinfo {author} {\bibfnamefont {M.}~\bibnamefont {Horodecki}},\ }\bibfield  {title} {\bibinfo {title} {Entanglement measures.},\ }\href@noop {} {\bibfield  {journal} {\bibinfo  {journal} {Quantum Inf. Comput.}\ }\textbf {\bibinfo {volume} {1}},\ \bibinfo {pages} {3} (\bibinfo {year} {2001})}\BibitemShut {NoStop}%
\bibitem [{\citenamefont {Koashi}\ and\ \citenamefont {Winter}(2004)}]{koashi2004monogamy}%
  \BibitemOpen
  \bibfield  {author} {\bibinfo {author} {\bibfnamefont {M.}~\bibnamefont {Koashi}}\ and\ \bibinfo {author} {\bibfnamefont {A.}~\bibnamefont {Winter}},\ }\bibfield  {title} {\bibinfo {title} {Monogamy of quantum entanglement and other correlations},\ }\href@noop {} {\bibfield  {journal} {\bibinfo  {journal} {Physical Review A}\ }\textbf {\bibinfo {volume} {69}},\ \bibinfo {pages} {022309} (\bibinfo {year} {2004})}\BibitemShut {NoStop}%
\bibitem [{\citenamefont {Vardeny}\ \emph {et~al.}(2013)\citenamefont {Vardeny}, \citenamefont {Nahata},\ and\ \citenamefont {Agrawal}}]{vardeny2013optics}%
  \BibitemOpen
  \bibfield  {author} {\bibinfo {author} {\bibfnamefont {Z.~V.}\ \bibnamefont {Vardeny}}, \bibinfo {author} {\bibfnamefont {A.}~\bibnamefont {Nahata}},\ and\ \bibinfo {author} {\bibfnamefont {A.}~\bibnamefont {Agrawal}},\ }\bibfield  {title} {\bibinfo {title} {Optics of photonic quasicrystals},\ }\href@noop {} {\bibfield  {journal} {\bibinfo  {journal} {Nature photonics}\ }\textbf {\bibinfo {volume} {7}},\ \bibinfo {pages} {177} (\bibinfo {year} {2013})}\BibitemShut {NoStop}%
\bibitem [{\citenamefont {Ozawa}\ and\ \citenamefont {Price}(2019)}]{ozawa2019topological}%
  \BibitemOpen
  \bibfield  {author} {\bibinfo {author} {\bibfnamefont {T.}~\bibnamefont {Ozawa}}\ and\ \bibinfo {author} {\bibfnamefont {H.~M.}\ \bibnamefont {Price}},\ }\bibfield  {title} {\bibinfo {title} {Topological quantum matter in synthetic dimensions},\ }\href@noop {} {\bibfield  {journal} {\bibinfo  {journal} {Nature Reviews Physics}\ }\textbf {\bibinfo {volume} {1}},\ \bibinfo {pages} {349} (\bibinfo {year} {2019})}\BibitemShut {NoStop}%
\bibitem [{\citenamefont {Li}\ \emph {et~al.}(2017)\citenamefont {Li}, \citenamefont {Li},\ and\ \citenamefont {Das~Sarma}}]{li2017mobility}%
  \BibitemOpen
  \bibfield  {author} {\bibinfo {author} {\bibfnamefont {X.}~\bibnamefont {Li}}, \bibinfo {author} {\bibfnamefont {X.}~\bibnamefont {Li}},\ and\ \bibinfo {author} {\bibfnamefont {S.}~\bibnamefont {Das~Sarma}},\ }\bibfield  {title} {\bibinfo {title} {Mobility edges in one-dimensional bichromatic incommensurate potentials},\ }\href@noop {} {\bibfield  {journal} {\bibinfo  {journal} {Physical Review B}\ }\textbf {\bibinfo {volume} {96}},\ \bibinfo {pages} {085119} (\bibinfo {year} {2017})}\BibitemShut {NoStop}%
\bibitem [{\citenamefont {Verresen}\ \emph {et~al.}(2021)\citenamefont {Verresen}, \citenamefont {Tantivasadakarn},\ and\ \citenamefont {Vishwanath}}]{verresen2021efficiently}%
  \BibitemOpen
  \bibfield  {author} {\bibinfo {author} {\bibfnamefont {R.}~\bibnamefont {Verresen}}, \bibinfo {author} {\bibfnamefont {N.}~\bibnamefont {Tantivasadakarn}},\ and\ \bibinfo {author} {\bibfnamefont {A.}~\bibnamefont {Vishwanath}},\ }\bibfield  {title} {\bibinfo {title} {Efficiently preparing schr$\backslash$" odinger's cat, fractons and non-abelian topological order in quantum devices},\ }\href@noop {} {\bibfield  {journal} {\bibinfo  {journal} {arXiv preprint arXiv:2112.03061}\ } (\bibinfo {year} {2021})}\BibitemShut {NoStop}%
\bibitem [{\citenamefont {Hogben}\ \emph {et~al.}(2011)\citenamefont {Hogben}, \citenamefont {Hore},\ and\ \citenamefont {Kuprov}}]{hogben2011multiple}%
  \BibitemOpen
  \bibfield  {author} {\bibinfo {author} {\bibfnamefont {H.}~\bibnamefont {Hogben}}, \bibinfo {author} {\bibfnamefont {P.}~\bibnamefont {Hore}},\ and\ \bibinfo {author} {\bibfnamefont {I.}~\bibnamefont {Kuprov}},\ }\bibfield  {title} {\bibinfo {title} {Multiple decoherence-free states in multi-spin systems},\ }\href@noop {} {\bibfield  {journal} {\bibinfo  {journal} {Journal of Magnetic Resonance}\ }\textbf {\bibinfo {volume} {211}},\ \bibinfo {pages} {217} (\bibinfo {year} {2011})}\BibitemShut {NoStop}%
\bibitem [{\citenamefont {Neschen}(1991)}]{neschen1991efficient}%
  \BibitemOpen
  \bibfield  {author} {\bibinfo {author} {\bibfnamefont {M.}~\bibnamefont {Neschen}},\ }\bibfield  {title} {\bibinfo {title} {An efficient multi-spin coding algorithm for neural networks},\ }\href@noop {} {\bibfield  {journal} {\bibinfo  {journal} {International Journal of Modern Physics C}\ }\textbf {\bibinfo {volume} {2}},\ \bibinfo {pages} {623} (\bibinfo {year} {1991})}\BibitemShut {NoStop}%
\bibitem [{\citenamefont {Higgins}\ \emph {et~al.}(2007)\citenamefont {Higgins}, \citenamefont {Berry}, \citenamefont {Bartlett}, \citenamefont {Wiseman},\ and\ \citenamefont {Pryde}}]{higgins2007entanglement}%
  \BibitemOpen
  \bibfield  {author} {\bibinfo {author} {\bibfnamefont {B.~L.}\ \bibnamefont {Higgins}}, \bibinfo {author} {\bibfnamefont {D.~W.}\ \bibnamefont {Berry}}, \bibinfo {author} {\bibfnamefont {S.~D.}\ \bibnamefont {Bartlett}}, \bibinfo {author} {\bibfnamefont {H.~M.}\ \bibnamefont {Wiseman}},\ and\ \bibinfo {author} {\bibfnamefont {G.~J.}\ \bibnamefont {Pryde}},\ }\bibfield  {title} {\bibinfo {title} {Entanglement-free heisenberg-limited phase estimation},\ }\href@noop {} {\bibfield  {journal} {\bibinfo  {journal} {Nature}\ }\textbf {\bibinfo {volume} {450}},\ \bibinfo {pages} {393} (\bibinfo {year} {2007})}\BibitemShut {NoStop}%
\bibitem [{\citenamefont {Cialdi}\ \emph {et~al.}(2011)\citenamefont {Cialdi}, \citenamefont {Brivio}, \citenamefont {Tesio},\ and\ \citenamefont {Paris}}]{cialdi2011programmable}%
  \BibitemOpen
  \bibfield  {author} {\bibinfo {author} {\bibfnamefont {S.}~\bibnamefont {Cialdi}}, \bibinfo {author} {\bibfnamefont {D.}~\bibnamefont {Brivio}}, \bibinfo {author} {\bibfnamefont {E.}~\bibnamefont {Tesio}},\ and\ \bibinfo {author} {\bibfnamefont {M.~G.}\ \bibnamefont {Paris}},\ }\bibfield  {title} {\bibinfo {title} {Programmable entanglement oscillations in a non-markovian channel},\ }\href@noop {} {\bibfield  {journal} {\bibinfo  {journal} {Physical Review A—Atomic, Molecular, and Optical Physics}\ }\textbf {\bibinfo {volume} {83}},\ \bibinfo {pages} {042308} (\bibinfo {year} {2011})}\BibitemShut {NoStop}%
\end{thebibliography}%

\end{document}